\documentclass{aa}  %>>> use this instead for A4 paper

\usepackage[varg]{txfonts}
\usepackage{amsmath,amsfonts,amssymb}
\usepackage{graphicx}
\usepackage[colorlinks=true, allcolors=blue]{hyperref}
\usepackage{multirow}
\usepackage{subfigure}
\usepackage{siunitx}
\usepackage{bm}
\usepackage{natbib}

\bibpunct{(}{)}{;}{a}{}{,}

\usepackage{color}

\begin{document}

\title{Numerical estimation of wavefront error breakdown in adaptive optics}

\author{F. Ferreira
\and{E. Gendron}
\and{G. Rousset}
\and{D. Gratadour}}
\institute{LESIA, Observatoire de Paris, Université PSL, CNRS, Sorbonne Université, Univ. Paris Diderot, Sorbonne Paris Cité, 5 place Jules Janssen, 92195 Meudon, France}

\date{Received 03/01/2018, Accepted 24/04/2018}

\abstract{}{Adaptive optics (AO) system performance is improved using post-processing techniques, such as point spread
function (PSF) deconvolution. The PSF estimation involves characterization of the different
wavefront (WF) error sources in the AO system. We propose a numerical error breakdown estimation tool that allows studying AO error source behavior such as their correlations. We also propose a new analytical model for anisoplanatism and bandwidth errors that were validated with the error breakdown estimation tool. This model is the first step for a complete AO residual error model that is expressed in deformable mirror space, leading to practical usage such as PSF reconstruction or turbulent parameters identification.}{We have developed in the computing platform for adaptive optics systems (COMPASS) code, which is an end-to-end simulation code using graphics processing units (GPU) acceleration, an estimation tool that provides a comprehensive error breakdown by the outputs of a single simulation run. We derive the various contributors from the end-to-end simulator at each iteration step: this method provides temporal buffers of each contributor. Then, we use this tool to validate a new model of anisoplanatism and bandwidth errors including their correlation. This model is based on a statistical approach that computes the error covariance matrices using structure functions.}{On a SPHERE-like system, the comparison between a PSF computed from the error breakdown with a PSF obtained from classical end-to-end simulation shows that the statistics convergence limits converge very well, with  a sub-percent difference in terms of Strehl ratio and ensquared energy at $5\frac{\lambda}{D}$ separation. A correlation analysis shows significant correlations between some contributors, especially WF measurement deviation error and bandwidth error due to centroid gain, and the well-known correlation between bandwidth and anisoplanatism errors is also retrieved. The model we propose for the two latter errors shows an SR and EE difference of about one percent compared to the end-to-end simulation, even if some approximations exist.}{} 

\titlerunning{Error breakdown estimation for adaptive optics systems}
\maketitle

\section{Introduction}
Optical aberrations that are due to turbulence have a huge impact on the image resolution of ground-based large telescopes at visible and infrared wavelengths. Without any compensation, the resolution in the visible is equivalent to a diffraction-limited image of a telescope with a few tens of centimeters in diameter. Adaptive optics (AO) systems have been developed for several years in astronomy \citep{Rousset1990} to compensate for these aberrations: a wavefront sensor (WFS) measures the wavefront deformation, and a deformable mirror (DM) is controlled in real time to flatten the wavefront for the scientific image.

However, the compensation by AO systems is not perfect: there is still a residual wavefront error that has variable impact on the image quality, depending on the system and the observing conditions. Post-processing techniques, such as point spread function (PSF) deconvolution algorithms, have been developed to improve the contrast on the final image \citep{Mugnier2004}. This approach requires estimating the PSF over the scientific field, which involves characterizing the error contributors of the AO system \citep{Veran1997, Harder2000, Jolissaint2004, Correia2011, Gendron2014, Martin2016}. Another approach is the use of end-to-end system simulation \citep{Gilles2012}. Estimating and distinguishing the various error contributors is a problem because of the propagation and filtering process of the errors in the AO loop \citep{Vidal2014,Juvenal2015, Martin2017}. 

End-to-end AO simulation tools use a Monte Carlo approach to provide an AO-corrected PSF. It requires several thousand iterations to achieve sufficient convergence on the PSF computation. At the ELT (extrememly large telescope) scale, it is particularly demanding in terms of computing power and data flow. Analytical approaches have been developed to provide analytical expression of the PSF \citep{Jolissaint2010, Gendron2014, Neichel2008}. If these models do not have convergence problems, they usually require several simplificating assumptions such as stationarity and statistical independence. 

In the first part of this paper, we present the development of a new estimation tool, called ROKET (error breakdown estimation tool), which provides a comprehensive error breakdown as an output of a single run of an end-to-end AO simulation \citep{Ferreira2016}. It is developed in the end-to-end graphics processing units (GPU)-based simulation tool COMPASS (computing platform for adaptive optics systems) \citep{Gratadour2014}. The novelty of ROKET is that all the error terms are determined frame by frame at each single iteration step of the simulation, instead of only assessing their statistical behavior. This is done by using known internal quantities that are computed by the simulation to estimate these errors
on the fly without interrupting the main AO loop. Thus, this tool provides an error breakdown with no other assumption than those that are made by the end-to-end models, which are usually weaker than the assumptions needed by analytical models. The goal of this tool is to quantify a detailed wavefront error breakdown in an AO system design by numerical simulations. In particular, statistical correlations between contributors can be evaluated using ROKET. The code can be use to study the behavior of error breakdown contributors and to validate or disprove strong assumptions made by an analytical model. 

In the second part of this paper, we propose an analytical model for estimating the anisoplanatism and bandwidth error terms, including their correlation. For instance, this model can be used in an a posteriori PSF reconstruction algorithm, complementary to a conventional analytical model as proposed by \cite{Veran1997} or \cite{Jolissaint2010}. This analytical model is dependent on a reduced number of parameters that might be identified on AO telemetry data and by turbulence profiling tools on site. The novelty of this model is that it does not rely on a Fourier analysis, but on the error covariance matrices directly expressed in the DM space. In addition, this model is a good candidate for parallel implementation, especially using GPU acceleration, leading to efficient computing that is suitable for ELT scale.

We have realized that some models that assume independence of classical AO error terms (bandwidth, anisoplanatism, etc.) sometimes fail to reproduce the PSF obtained with an end-to-end model. Therefore, it is quite understandable that mismatches of a reconstructed PSF on a real system exist. If progress has to be made in this field, it has to first address the mismatches between reconstructed PSF on simulation data. Our study intends to identify at the simplest and lowest simulation level how the errors are mixed up and what a PSF reconstruction model should take into account to reproduce end-to-end simulated data with a reliable level of accuracy. Our work in this paper intends to pave the way for a future work on PSF reconstruction that is based on telemetry data of real systems, but it remains preliminary to this future
work.

After we describe the error breakdown estimation tool in Section~\ref{sec::ROKET}, we discuss the error correlations that could be observed in Section~\ref{sec:CorrelationAnalysis}. In Section~\ref{sec::Model} we propose a pseudo-analytical model for bandwidth and anisoplanatism errors. We also discuss the results obtained with this model and its limitations in this section. The computing performance is summarized in Section~\ref{sec::computing_performance}.

%%%%%%%%%%%%%%%%%%%%%%%%%%%%%%%%%%%%%%%%%%%%%%%%%%%%
%%%%%%                                                                  AO LOOP ERROR                                                                                      %%%%
%%%%%%%%%%%%%%%%%%%%%%%%%%%%%%%%%%%%%%%%%%%%%%%%%%%%

\section{Error breakdown estimation tool}
\label{sec::ROKET}
In this section, we present our error breakdown estimation tool. After recalling the equations that have been derived in \cite{Ferreira2016}, we validate its outputs against a classical run of COMPASS. In order to make it as simple as possible in a first step, we consider an AO system with a single wavefront sensor and a single deformable mirror coupled to a tip-tilt mirror both conjugated to ground, also called single conjugated adaptive optics (SCAO) system.

\subsection{AO loop error}
A wavefront sensor requires a bright guide star (GS): the observed scientific object is sometimes too faint to allow the wavefront measurement. In that case, a guide star is chosen that lies as close as possible to the scientific object in order to limit the anisoplanatism error \citep{Fried1982}. The residual phase $\Phi_{\epsilon}(\mathbf{x},\bm{\theta})$ seen by the WFS is the turbulent phase in the GS direction $\Phi(\mathbf{x},\bm{\theta})$, corrected by the DM, which adds a phase $\Phi_{DM}(\mathbf{x})$:
\begin{equation}
    \label{eq:residual}
    \Phi_{\epsilon}(\mathbf{x},\bm{\theta}) = \Phi(\mathbf{x},\bm{\theta}) + \Phi_{DM}(\mathbf{x}) 
\end{equation}
with $\bm{\theta}$ the GS direction. Hence, we note with $\Phi(\mathbf{x},0)$ the turbulent phase in the scientific direction.
These quantities can be expressed on a basis $\mathcal{B}$ of the DM:
\begin{eqnarray}
    \Phi_{\epsilon}(\mathbf{x},\bm{\theta}) &=& \sum_{i=1}^N \epsilon_{i,\bm{\theta}} \, \mathcal{B}_i(\mathbf{x}) + \Phi_{\perp}(\mathbf{x},\bm{\theta})\\
    \Phi(\mathbf{x},\bm{\theta}) &=& \sum_{i=1}^N a_{i,\bm{\theta}} \, \mathcal{B}_i(\mathbf{x}) + \Phi_{\perp}(\mathbf{x},\bm{\theta}) \\
    \label{eq:DMmodalDecompo}
    \Phi_{DM}(\mathbf{x}) &=& \sum_{i=1}^N v_i \, \mathcal{B}_i(\mathbf{x})
,\end{eqnarray}
where $N$ is the number of commanded modes and $\Phi_{\perp}(\mathbf{x},\bm{\theta})$ is the component of $\Phi(\mathbf{x},\bm{\theta})$ that is orthogonal to the DM space. We note with $\bm{\epsilon}$, $\mathbf{a,}$ and $\mathbf{v}$ the vectors composed of the coefficients $\epsilon_{i,\theta}$, $a_{i,\theta,}$ and $v_i,$ respectively. For clarity, we simplify the notation for the scientific direction:
\begin{eqnarray}
    \bm{\epsilon} &=& \bm{\epsilon}_{\theta=0} \\
    \mathbf{a} &=& \mathbf{a}_{\theta=0}
.\end{eqnarray}

Then, assuming here a linear wavefront sensor, a system with a one-frame delay, and using Eqs.~(\ref{eq:residual}) to~(\ref{eq:DMmodalDecompo}), the measurement vector $\mathbf{w}_k$ of the sensor at the iteration $k$ of the AO loop can be written as 
\begin{equation}
\label{eq::WFS_measure}
\mathbf{w}_k = D \mathbf{a_{k,\theta}} + D \mathbf{v_{k-1}} + \mathbf{r}_k + \mathbf{n}_k + \mathbf{u}_k
,\end{equation}
where $D$ is he interaction matrix, and $\mathbf{n}_k$ is the noise contribution on the measurement vector. $\mathbf{u}_k$ covers anything that is not noise and that deviates from the theoretical model of ideal wavefront slope measurements, such as centroid gain and truncation effect. We call this term the wavefront measurement deviation. $\mathbf{r}_k$ is the aliasing contribution in the GS direction \citep{Veran1997, Rigaut1998}, 
\begin{equation}
    \mathbf{r}_k = \mathcal{M}(\Phi_{\perp}^k(\mathbf{x},\bm{\theta}))
,\end{equation}
with $\mathcal{M}$ the linear operator that describes the ideal wavefront slope measurement from any input phase.

As the system works in closed loop, we assume that the temporal command law used in real time is an integrator with a gain $g$:
\begin{equation}
\label{eq:integwithgain}
\mathbf{v}_k = \mathbf{v}_{k-1} - g R \mathbf{w}_k
,\end{equation}
where $R$ is the command matrix, which classically computed as the pseudo-inverse of the interaction matrix $D$.

Then, if we express the residual phase in the scientific direction on the DM basis using Eq.~(\ref{eq:residual}) considering a one-frame delay, we obtain a definition of the error without the fitting term:
\begin{equation}
\label{eq:errordef}
\bm{\epsilon}_{k} = \mathbf{a}_{k} + \mathbf{v}_{k-1}
.\end{equation}

Finally, we can derive the error at the iteration $k$ from Eqs.~(\ref{eq::WFS_measure}),~(\ref{eq:integwithgain}), and~(\ref{eq:errordef}):
\begin{multline}
\label{eq:commanderror}
\bm{\epsilon}_{k} = (1 - g R D) \bm{\epsilon}_{k-1} + (\mathbf{a}_{k} - \mathbf{a}_{k-1})+ \\ g R D(\mathbf{a}_{k-1}-\mathbf{a}_{k-1,\bm{\theta}}) 
-g R \mathbf{r}_{k-1}- g R \mathbf{n}_{k-1} - g R \mathbf{u}_{k-1}
\end{multline}

%%%%%%%%%%%%%%%%%%%%%%%%%%%%%%%%%%%%%%%%%%%%%%%%%%%%
%%%%%%                                                                  ERROR BREAKDOWN CONTRIBUTORS                          %%%%
%%%%%%%%%%%%%%%%%%%%%%%%%%%%%%%%%%%%%%%%%%%%%%%%%%%%
\subsection{Error breakdown contributors}
\label{section:contributors}

Writing Eq.(~\ref{eq:commanderror}) in this way allows us to reveal five contributors to $\bm{\epsilon}_k$:
\begin{equation}
\label{eq:commanderror_simple}
\bm{\epsilon}_{k} = \bm{\beta}_k + \bm{\tau}_k + \bm{\rho}_k + \bm{\eta}_k + \bm{\mu}_k
,\end{equation}
with
\begin{eqnarray}
    \label{eq:bandwidth}
    \bm{\beta}_k &=& (1 - g R D) \bm{\beta}_{k-1} + (\bm{a}_k - \bm{a}_{k-1}) \\
    \label{eq:aniso}
    \bm{\tau}_k &=& (1 - g R D) \bm{\tau}_{k-1} + g  R D (\bm{a}_{k-1} - \bm{a}_{k-1,\theta}) \\
    \label{eq:aliasing}
    \bm{\rho}_k &=& (1 - g R D) \bm{\rho}_{k-1} - g R \bm{r}_{k-1} \\
    \label{eq:noise}
    \bm{\eta}_k &=& (1 - g R D) \bm{\eta}_{k-1} - g R \bm{n}_{k-1}\\
    \label{eq:nlin}
    \bm{\mu}_k &=& (1 - g R D) \bm{\mu}_{k-1} - g R \bm{u}_{k-1}
.\end{eqnarray}

\subsubsection{Bandwidth error $\mathbf{\beta}_k$}
$\bm{\beta}_k$ is the temporal error that is due to the delay between the time when the command is computed and the time when this command is applied on the DM. During the computation, the turbulent phase is evolving, whereas the DM shape does not. This can be interpreted as the difference between the actual turbulent phase at time $k$ and the command that is applied at the same time. 
This term only involves the phase in the direction of the scientific object.

\subsubsection{Anisoplanatism error $\mathbf{\tau}_k$}
$\bm{\tau}_k$ is an anisoplanatism error that is due to the wavefront difference between the scientific direction and the analysis direction. 
This anisoplanatism error is estimated in the DM space, and is
filtered by the temporal response through the AO loop.

\subsubsection{Aliasing error $\mathbf{\rho}_k$}
$\bm{\rho}_k$ is the aliasing term: the high frequencies of a
turbulent phase are misinterpreted by the WFS and result in a non-null measurement, reconstructed and compensated for by the AO system, introducing an aliased phase component. 

\subsubsection{Noise measurement error $\mathbf{\eta}_k$}
The noise on the WFS measurements is the term $\bm{\eta}_k$, which results from both read-out noise and photon noise on the WFS image. Only the low-frequency part of this noise is injected in the command. 

\subsubsection{WF measurement deviation error $\mathbf{\mu}_k$}
$\bm{\mu}_k$ describes anything that deviates from an ideal wavefront slope measurement. An example of it is the truncation effect, which occurs when the field of view of subapertures of a Shack-Hartmann is too small compared to the spot size. Another example is the undersampling error that is due to the extended size of the pixel of a Shack-Hartmann. It can be estimated numerically by the difference between the WFS measurements without noise and the measurements obtained by directly computing the phase gradient of the wavefront.

\subsubsection{Fitting and filtered mode error $\Phi_\perp$}
The residual phase $\Phi_{\perp}$ includes a term from the filtered modes and a fitting term. These two contributions are evaluated separately.

The filtered mode error is due to the filtering process used to invert the interaction matrix. It can be estimated easily if the subspace of the modes commanded through the system control matrix is orthogonal to the subspace of filtered ones.

The fitting term is derived as the residual phase that is left after the projection onto the DM basis. As this phase is orthogonal to the DM space, it is not expressed on any DM basis. In order to estimate its impact on the Strehl ratio (SR), we compute its
spatial variance at each iteration.

\subsection{Modal basis}
\label{modalBasis}
In order to exploit the error breakdown estimation for a PSF reconstruction as an example, it is more convenient to express the different contributors as a phase variance spread over a modal basis. Such modal variances can be derived from the error terms obtained and the knowledge of the DM basis or influence function used to control it. To handle it, we built a modal basis $\mathcal{B}_{tt}$ from the influence functions
with the following properties:
\begin{itemize}
\item The modes are orthonormal.
\item In particular, the subspace of the modes commanded through the system control matrix is orthogonal (same scalar product as above) to the subspace of filtered modes.
\item The modes are orthogonal to piston and tip-tilt modes.
\item Two of these modes are pure tip-and-tilt modes.
\end{itemize}
%This property is essential to conserve the phase variance in both influence function and modal spaces. In this basis, we are able to separate the filtered modes from the commanded ones and to estimate the phase variance due to this contribution.
See Appendix~\ref{app::modal_basis} for details on the computation of this modal basis.

\subsection{Validation against end-to-end simulation}
\label{sec::results}
As ROKET provides the temporal buffers of each contributor estimated in the DM command space, we are able to compute any covariance matrix between two error terms and the full residual error covariance matrix as

\begin{equation}
    \label{eq::coverrtot}
\langle \bm{\epsilon} \bm{\epsilon}^t \rangle = \langle \left( \bm{\beta} + \bm{\tau} + \bm{\rho} + \bm{\eta} + \bm{\mu}  \right) \left( \bm{\beta} + \bm{\tau} + \bm{\rho} + \bm{\eta} + \bm{\mu} \right)^t \rangle
.\end{equation}
From this matrix, we are able to reconstruct the AO PSF using algorithms developed by \citet{Veran1997} or \citet{Gendron2006}. These algorithms provide an estimate of the optical transfer function by computing a structure function from the AO residual error covariance matrix $\langle \bm{\epsilon} \bm{\epsilon}^t \rangle$. In this paper, we use the algorithm proposed by \citet{Gendron2006}, which uses the so-called $V_{ii}$ functions that can be computed on the fly. This makes it more convenient for a GPU implementation because only a limited amount of memory is available on the device.

Then, we now have two ways of producing a PSF after a simulation run.
The built-in way is naturally produced by COMPASS itself as a simulation output. This is noted PSF$_C$ and is regarded as the reference PSF.
The other way is computed from ROKET data using a PSF reconstruction algorithm.

We propose to compare the two PSFs and discuss their difference.
We have chosen to compare them in terms of maximum value (or SR) and ensquared energy (EE) at a half-width of $5 \frac{\lambda}{D}$.
The importance of the error contributors and their possible correlations are then studied in detail in Sect.~\ref{sec:CorrelationAnalysis}.

\subsubsection{Simulation parameters}
\label{sec:cases}
We consider a telescope with a diameter of 8m without central obstruction. The parameters used in all the simulations are listed in Table~\ref{table:params}. It corresponds to a SPHERE-like system \citep{Beuzit2008}.
\begin{table*}[!htb]
\centering
\begin{tabular}{l l l l}
\hline
\hline
\multicolumn{2}{c}{\textbf{Telescope parameters}} & \multicolumn{2}{c}{\textbf{Target parameters}}\\
\hline
Diameter & 8 m & Wavelength $\lambda_t$ & 1.65 $\mu m$ \\
\hline
\multicolumn{2}{c}{\textbf{Atmospheric parameters}} & \multicolumn{2}{c}{\textbf{WFS parameters}}\\
\hline
Number of layers & 12 & Number of subapertures & $40 \times 40$\\
$r_0$ (500 nm) & 0.16 m & Wavelength $\lambda_{wfs}$ & 0.5 $\mu m$ \\
$L_0$& 100 m & Number of pixels per subap. & 6 \\
& & Pixel size & 0.5" \\
& & Photons per subap. & 760 \\
 &  & Readout noise & 3 $e^-$  \\
 & & Centroiding method & Classical CoG\\
 &  & Guide star coordinates in FoV & (5'',0'')\\

\hline
\multicolumn{2}{c}{\textbf{AO parameters}} & \multicolumn{2}{c}{\textbf{DM parameters}} \\
\hline
Loop frequency & 500 Hz & Number of DM actuators & 41 $\times$ 41 \\
Command law & Integrator & 1 tip-tilt mirror & \\
Loop gain & 0.3 & Conjugation altitude & 0 m (pupil)\\
Delay & 1 frame & & \\
Frames & 20\,000 & &\\
\hline
\end{tabular}
\caption{\label{table:params} Simulation parameters}
\end{table*}
The parameters regarding the atmospheric conditions are detailed in Table~\ref{table::12-layers}.
The simulations was run over $20\,000$ iterations, which is  equivalent to 40 seconds of observation. The sky coordinates and directions are defined in a reference frame where the center is the science target and the X-axis is oriented toward the WFS guide star.
\begin{table}
\centering
\begin{tabular}{l l l l}
\hline
\hline
Alt. [m] & Wind speed [m/s] & Wind dir. [deg] & Frac. of $C_n^2$\\ %$r_0$ \\
\hline
0 & 13 & 345 & 0.261 \\
100 & 17 & 68 & 0.138 \\
200 & 5 & 245 & 0.081 \\
300 & 17 & 199 & 0.064 \\
900 & 10 & 181 & 0.105 \\
1800 & 10 & 94 & 0.096 \\
4500 & 8 & 152 & 0.085 \\
7100 & 6 & 185 & 0.053 \\
11\,000 & 14 & 265 & 0.048 \\
12\,800 & 9 & 116 & 0.037 \\
14\,500 & 8 & 6 & 0.021 \\ 
16\,500 & 17 & 272 & 0.011 \\
\hline
\end{tabular}
\caption{\label{table::12-layers} 12-layer turbulent profile}
\end{table}

%For this study, the system geometry has been fixed and we ran simulations by varying the loop gain, the wind direction and speed for a total of 60 simulations. Each case was simulated over 20000 iterations equivalent to 40 seconds of observation.

\subsubsection{Results}
\label{sec::multi-layers}
We can directly deduce from the ROKET outputs the conventionnal error breakdown as shown in Table~\ref{table::breakdown}. Cross-terms are not included in this table and are detailed in Section~\ref{sec:CorrelationAnalysis}.

\begin{table}[!htb]
\centering
\begin{tabular}{l c}
\hline
\hline
\textbf{Contributors} & \textbf{$\mathbf{\sigma}$ [nm rms]} \\
\hline
Bandwidth & 80 \\
Anisoplanatism & 109 \\
Aliasing & 31 \\
Noise & 11 \\
WF measurement deviation & 7.5 \\
Fitting & 54 \\
Filtered modes & 6 \\
\hline
\end{tabular}
\caption{\label{table::breakdown} Error breakdown returned by ROKET for the test case}
\end{table}

The error breakdwon obtained is dominated by the anisoplanatism and bandwidth terms, and it is compliant with common approximations used to estimate the contributors, such as the fitting error estimated from \citep{Hudgin1977} using $\sigma_{fit}^2 \approx 0.23 \left( \frac{d}{r_0} \right)^{\frac{5}{3}} $ (with $d$ the inter actuator distance), leading here to 46~nm~rms. The variance of aliasing, set to 33\% of the fitting \citep{Rigaut1998}, gives 31~nm~rms.

\begin{figure}%[!htbp]
\resizebox{\hsize}{!}{\includegraphics[trim={8cm 0 0 0},clip]{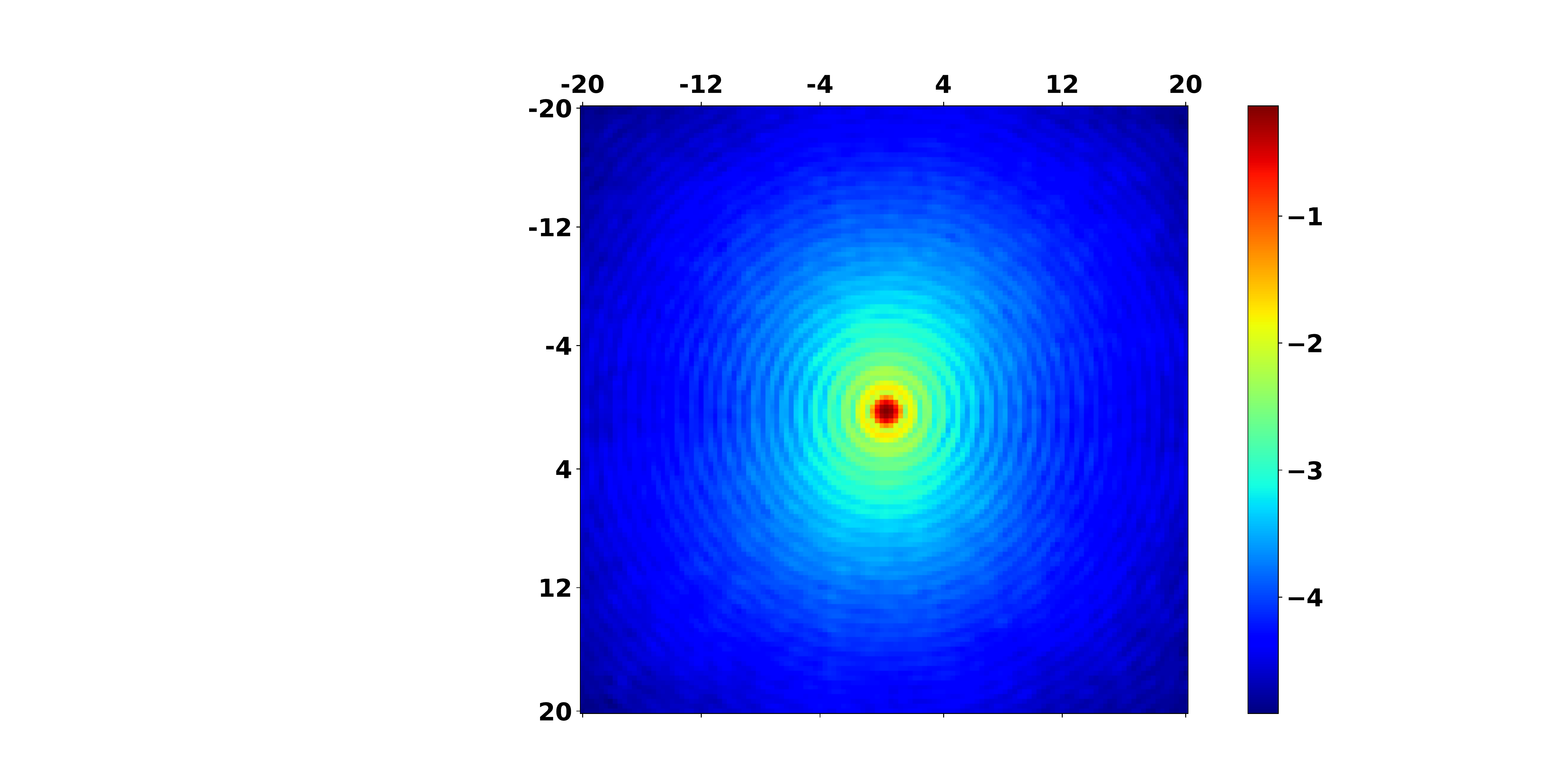}}
\resizebox{\hsize}{!}{\includegraphics[trim={8cm 0 0 0},clip]{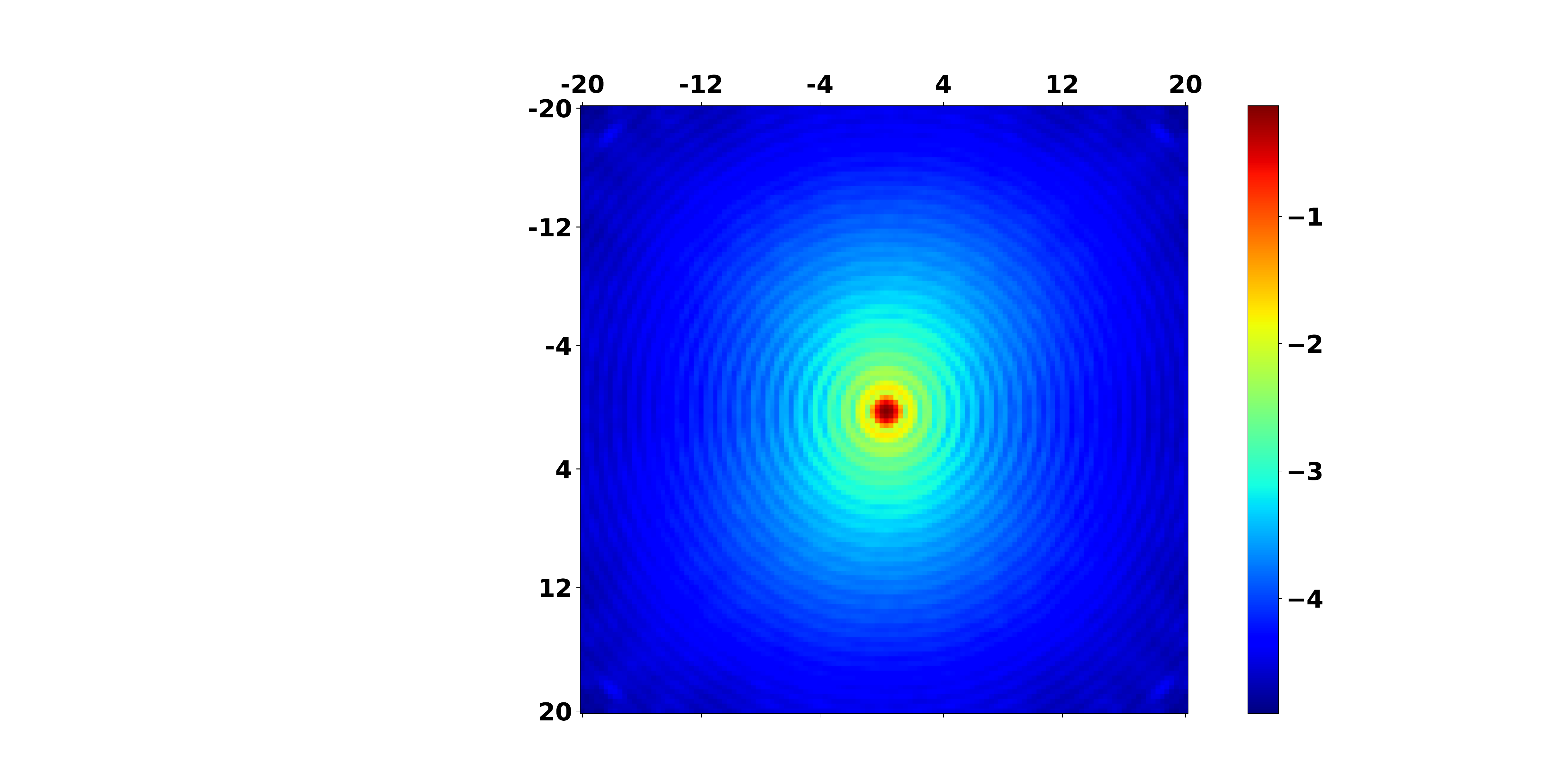}}
\resizebox{\hsize}{!}{\includegraphics[trim={8cm 0 0 0},clip]{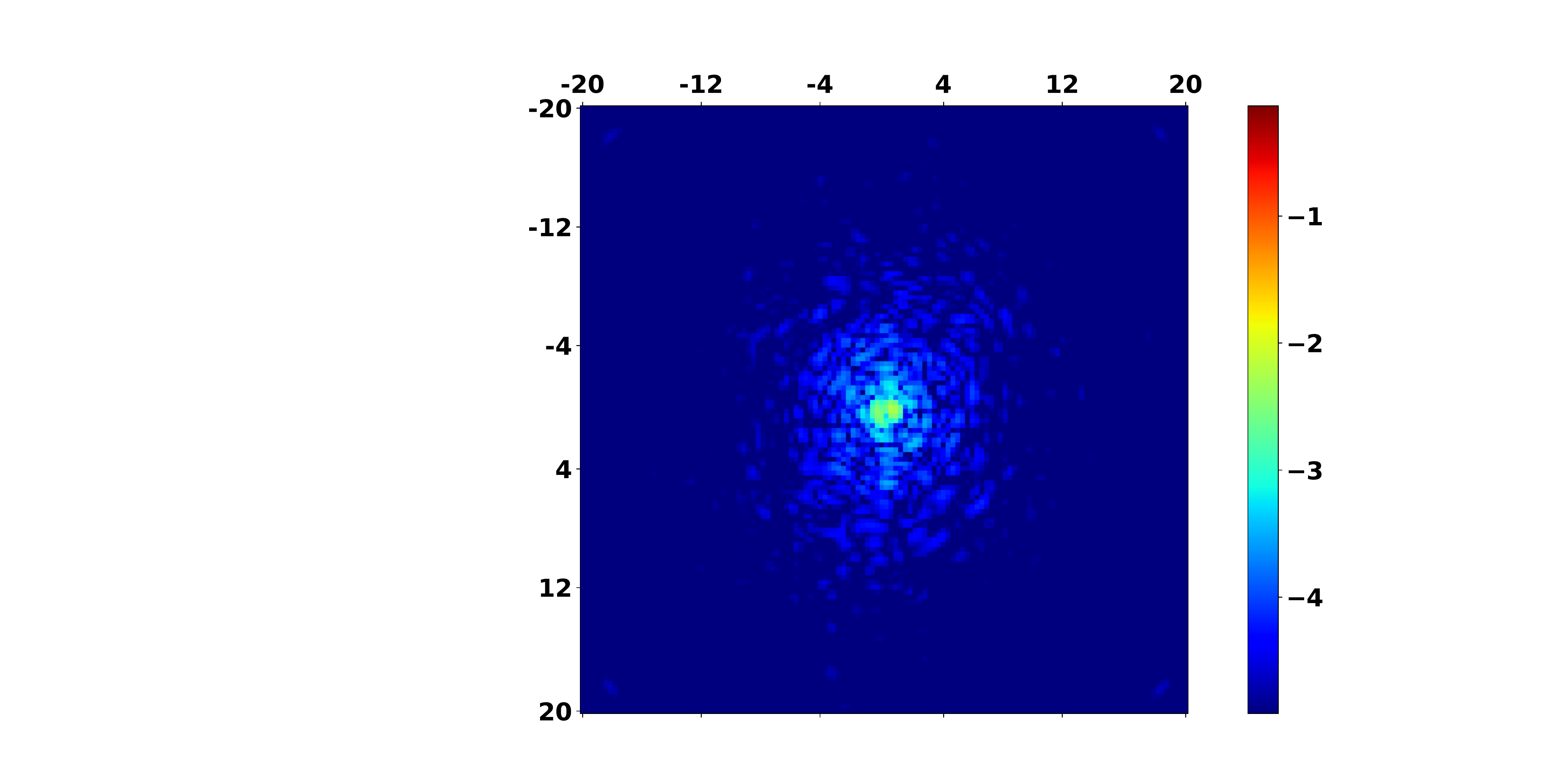}}
\caption{\label{fig::PSF_validation} PSFs in log scale. Axes are expressed in units of $\frac{\lambda}{D}$. Top: PSF$_C$ simulated by COMPASS. Middle: PSF$_R$ reconstructed from ROKET buffers. Bottom: $\| PSF_C - PSF_R \|.$}
\end{figure}

\begin{figure}%[!htbp]
\resizebox{\hsize}{!}{\includegraphics[width=14cm]{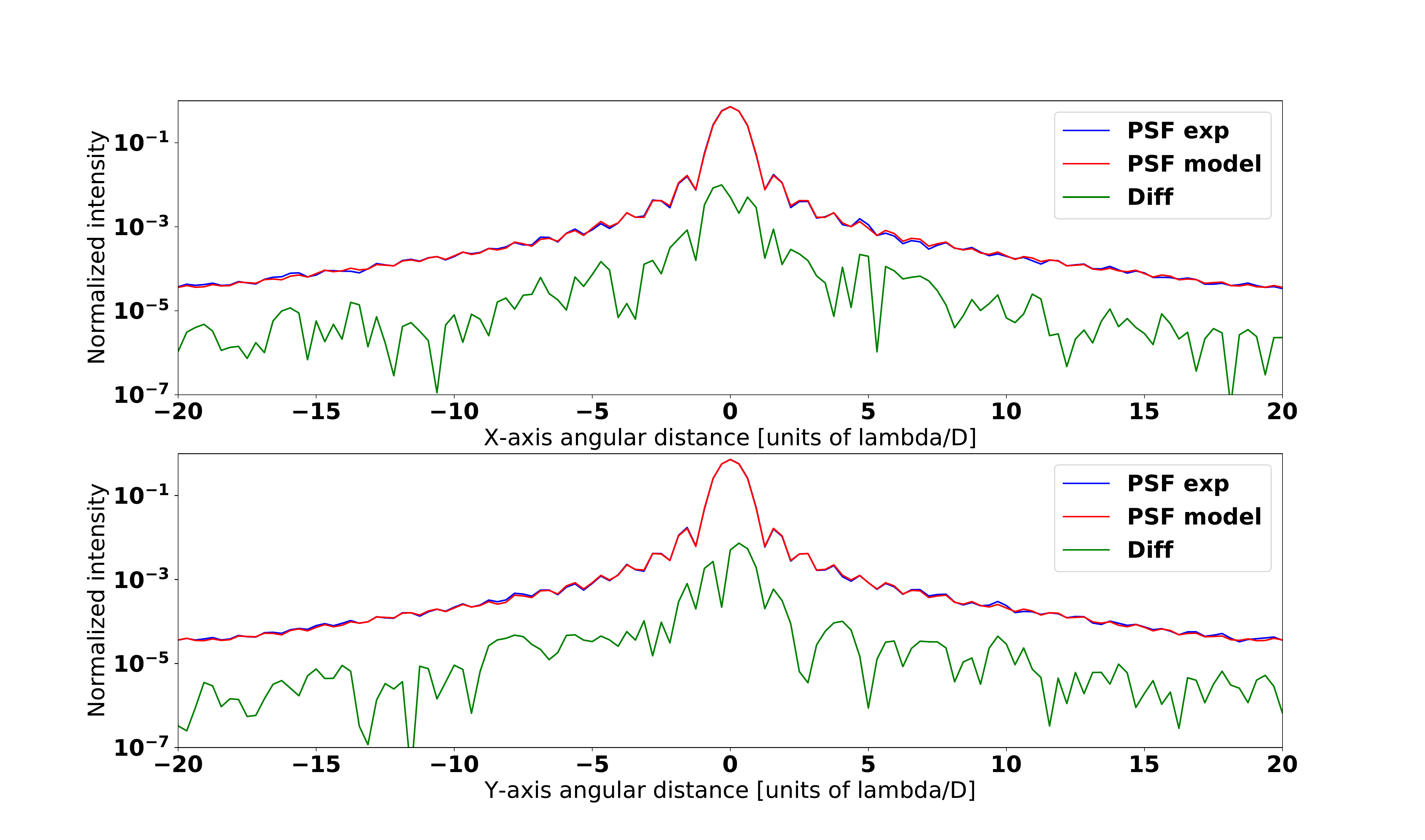}}
\caption{\label{fig::PSF_cuts_validation} Cuts of PSF$_C$ in red and PSF$_R$ in blue. The green curve is $\| PSF_C - PSF_R \|$. Top: Cut along X-axis. Bottom: Cuts along Y-axis.}
\end{figure}

To validate this error breakdown, we reconstructed the PSF$_R$ by directly computing the residual error covariance matrix from the sum of the error buffers returned by ROKET just as in Eq.~(\ref{eq::coverrtot}), and we compared it to the PSF$_C$ simulated by COMPASS. Figure~\ref{fig::PSF_validation} shows these two PSFs and their absolute difference in log scale. Cuts along the X- and Y-axes are shown in Fig.~\ref{fig::PSF_cuts_validation}.

We note the perfect agreement between the two PSFs, with an estimated SR of 73.6\% by ROKET compared to 73.8\% computed by COMPASS, and an EE of 83.2\% compared to 83.3\%, respectively. This demonstrates that the estimation of ROKET is accurate, as we can reliably retrieve the PSF from it. Of course, the PSF$_C$ exhibits a speckled aspect because the simulation does not converge, while the PSF$_R$ is built from an averaged phase structure function \citep{Veran1997} and thus appears smooth and symmetric.

%%%%%%%%%%%%%%%%%%%%%%%%%%%%%%%%%%%%%%%%%%%%%%%%%%%%%%%
%%%%%%                  Correlation analysis               %%%%
%%%%%%%%%%%%%%%%%%%%%%%%%%%%%%%%%%%%%%%%%%%%%%%%%%%%%%%

\section{Correlation analysis}
\label{sec:CorrelationAnalysis}
This section takes advantage of ROKET capacities to highlight the correlations that exist between some of the error breakdown contributors. We also study the conditions that lead to these correlations.

\subsection{Correlation between error terms}

Developing  Eq.~\ref{eq::coverrtot} leads to 5 main terms that are the covariance matrices of each error breakdown contributor, and to 20 cross-terms that are often neglected in many studies \citep{Martin2016, Jolissaint2010, Vidal2014}.
%Usually, as we cannot estimate the error contributors as ROKET. Instead, we estimate or measure each contributor separately and then, to obtain the global error, we suppose that each contributor is statiscally independent from the other. It is equivalent to consider that covariances between contributors are negligible.
A statistical independence between the error contributors is usually assumed, so that Eq.~(\ref{eq::coverrtot}) simplifies to
\begin{equation}
\label{eq::coverrind}
\langle \bm{\epsilon} \bm{\epsilon}^t \rangle = \langle \bm{\beta} \bm{\beta}^t \rangle + \langle \bm{\tau} \bm{\tau}^t \rangle + \langle \bm{\eta} \bm{\eta}^t \rangle + \langle \bm{\mu} \bm{\mu}^t \rangle + \langle \bm{\rho} \bm{\rho}^t \rangle
,\end{equation}
which is, in many cases, a meaningful hypothesis.

However, we cannot infer with certainty that this assumption is always valid. Obviously, some contributors of the error breakdown are effectively independent (such as noise with respect to turbulent terms, e.g.), but we have no guarantee that turbulence-related terms are independent among themselves.
This assumption could have an impact in the PSF estimation as it directly affects the error covariance matrix.
With ROKET, we will be able to assess the validity of assuming
that the error term is statistically independent, and determine what to do if this is not true.

\begin{figure}
    \centering
    \resizebox{\hsize}{!}{\subfigure{\includegraphics[trim={0 2cm 0 0},clip]{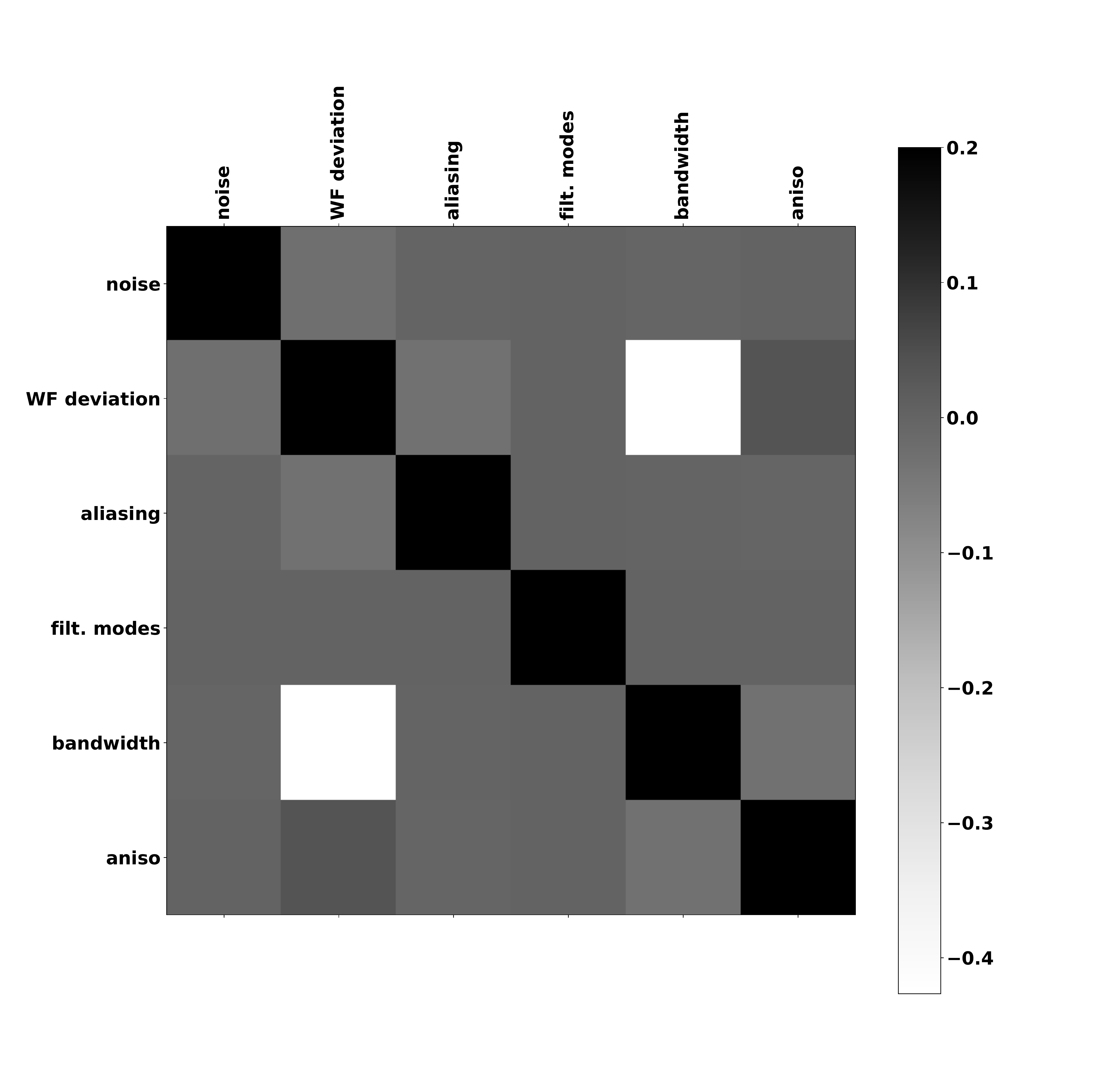}}}
    \caption{\label{fig::cov_cor_nlin} Correlation matrix between the error contributors. The color scale has been changed to highlight small cross-correlations. The diagonal is still equal to 1.}
    \end{figure}

We are able to compute all covariance terms, including cross-terms, from the temporal buffers of each error contributor returned by ROKET. Figure~\ref{fig::cov_cor_nlin} shows the correlation coefficients obtained between all error contributors for the simulated case described in Section~\ref{sec:cases}. They are displayed in matrix form, with lines and rows associated with a given contributor. Thus, for two contributors $x$ and $y$, the value of the cell that intersects line $x$ and column $y$ is the correlation coefficient between these terms.
We note a strong anticorrelation of -0.43 between the WF measurement deviation and bandwidth. Other non-null correlations are a WF measurement deviation with aliasing (-0.03), anisoplanatism (0.03), noise (-0.02), and also bandwidth and anisoplanatism (0.03). The WF measurement deviation affects the measurements of the low-order modes, so that it is naturally correlated to the temporal and the aliasing error, which are both made of low orders. ROKET shows a correlation between WF measurement deviation and noise. The Shack-Hartmann image profile affects the centroiding noise $\bm{n}_k$ and also the centroid gain, which is part of the deviation term. Consequently, it is impossible to argue that these terms cannot be correlated, as they both come from the centroiding process of the image.

\subsection{WF measurement deviation and bandwidth correlation}
\label{ssec:correlationStudy}

The correlation between deviation and bandwidth is first due to a centroid gain that can be modeled by a multiplicative scalar factor $\gamma$ on the WFS measurement. By construction of ROKET, the effect of the centroid gain is counted as part of $\mathbf{u}_k$ --an odd behavior.
On an other hand, the effect of a centroid gain is often regarded as the equivalent to an alteration of the loop gain and is expected to primarily affect the bandwidth term, and to a lesser extent the propagation of noise, aliasing, and other terms \citep{Veran2000}. It would make more sense to identify this centroid gain as such to establish a new error breakdown. We have modified the wavefront measurement models described in \citet{Ferreira2016} for this purpose. For the sake of clarity, the new equations are presented in Appendix~\ref{app::centroid_gain}, Eq.~(\ref{eq:newCentroidGain}). We highlight one point: the summation of the five equations obtained in (\ref{eq:newCentroidGain}) will always result in the same $\bm{\epsilon}_k$, by definition. If the decomposition is affected by $\gamma$, the total remains unchanged because the AO system is not retuned with respect to $\gamma$. As a result, different decompositions using different gamma will lead to exactly the same PSF. The $\gamma$ parameter only changes the breakdown by transferring variances and covariances from one item to another. Now, using the particular value of $\gamma$ that matches the system centroid gain has some particular property: it causes
the covariances to vanish.

Of course, a new simulation run needs to be performed because the centroid gain $\gamma$ also needs to be applied to the phase gradient WFS model. To compute the correct value of $\gamma$, ROKET performs a linear regression between the WFS measurement without noise and the phase gradient model measurement, in a first run. The value of $\gamma$ is then estimated as the average of the regression coefficients over the iterations. Finally, we need to launch a new ROKET simulation run with this $\gamma$ value in order to compute the error breakdown as given in Appendix~\ref{app::centroid_gain}.

\begin{figure}
\centering
\resizebox{\hsize}{!}{\subfigure{\includegraphics[trim={-2cm 2cm 0 0},clip]{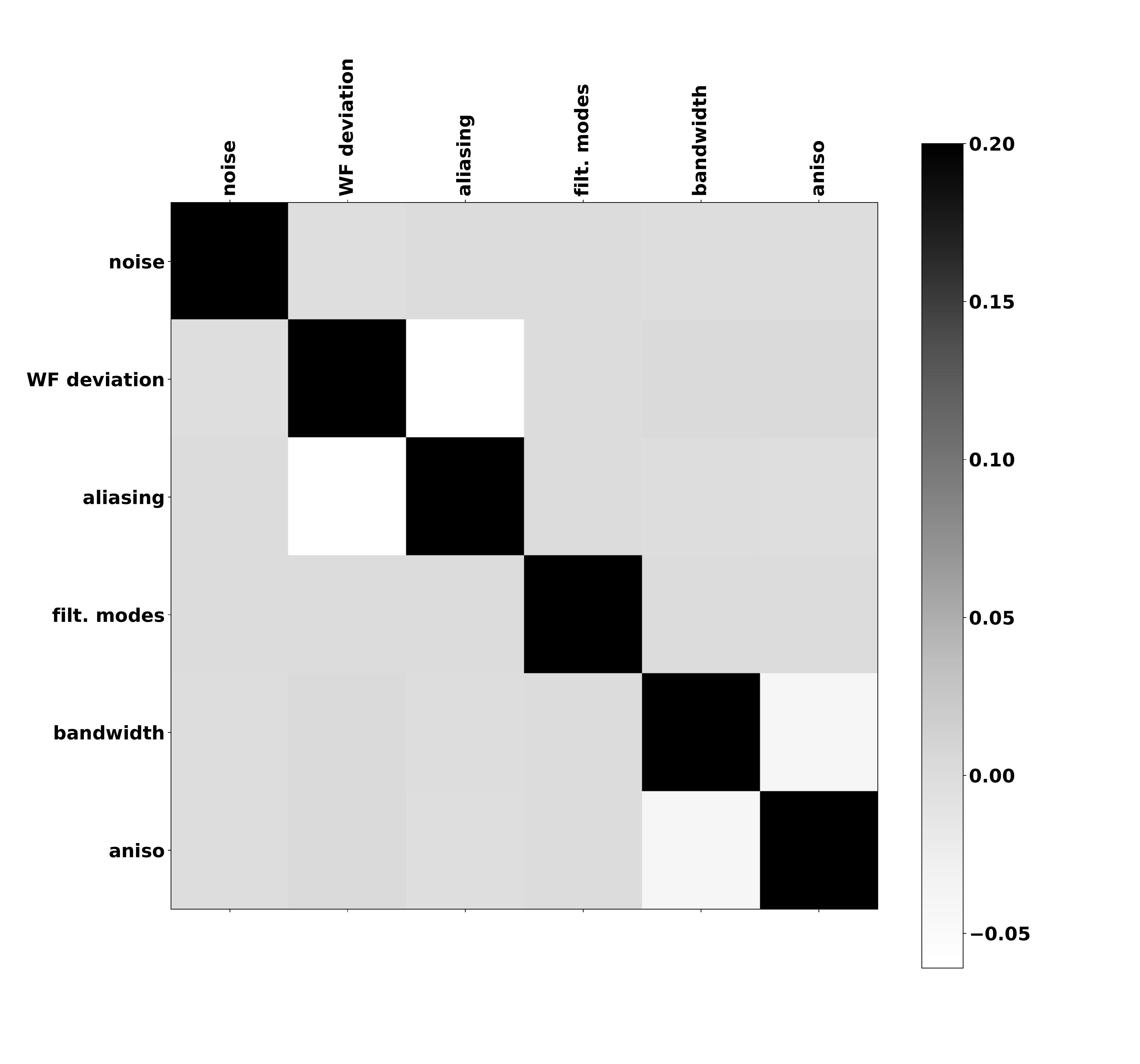}}}
\caption{\label{fig::cov_cor_nlin_cgain} Correlation matrix between the error contributors by taking into account the centroid gain. The color scale has been changed to highlight the vanishing of the deviation and bandwidth cross-correlation. The diagonal is still equal to 1.}
\end{figure}

Figure~\ref{fig::cov_cor_nlin_cgain} shows the correlation coefficients obtained with these equations. We note that the WF measurement deviation term is significantly reduced and the correlation between WF measurement deviation and bandwidth has been zeroed. We also note the disappearance of the correlation between WF measurement deviation and noise, which was then effectively due to the existence of this centroid gain.
A correlation remains with aliasing.
\begin{table}
    \centering
    \begin{tabular}{l l l l l l l}
    \hline
    \hline
    $\bm{\gamma}$ & $\bm{\sigma^2_\eta}$ & $\bm{\sigma^2_\mu}$ & $\bm{\sigma^2_\rho}$ & $\bm{\sigma^2_\beta}$ & $\bm{\sigma^2_{\tau}}$ & $\bm{\sigma^2_{\mu \beta}}$\\
    \hline
    1 & 0.114 & 0.057 & 0.948 & 6.43 & 11.8 & -0.258 \\
    0.95 & 0.109 & 0.036 & 0.908 & 5.93 & 12.0 & 0 \\
    \hline
    \end{tabular}
    \caption{\label{table::varcov} Variances and covariances obtained with $\gamma=1$ (i.e., without taking the centroid gain
into account) and with $\gamma = 0.95$ (value computed after the first run of ROKET). $\sigma^2_\eta$ is thenoise variance, $\sigma^2_\mu$ is the WF measurement deviation variance, $\sigma^2_\rho$ is the aliasing variance, $\sigma^2_\beta$ is the bandwidth variance, $\sigma^2_\tau$ is the anisoplanatism variance, and $\sigma^2_{\mu \beta}$ is the covariance between the WF measurement deviation and bandwidth. All are expressed in $10^{-3}\mu m^2$}
\end{table}
Introducing this centroid gain $\gamma$ in the equations has moved variances and covariances around, as shown in Table~\ref{table::varcov}. It is worth noticing that the error breakdown returned by ROKET remains valid in any case, since the sum of the time buffers remains unchanged. Similarly, the sum of all elements of the covariance tables in Figs.~\ref{fig::cov_cor_nlin} and~\ref{fig::cov_cor_nlin_cgain} leads to the same total variance, regardless of the value of $\gamma$. Taking $\gamma$ into account is only a modification of the wavefront measurement models, leading to a new error breakdown.

\subsection{Anisoplanatism and bandwidth correlation}
\label{sec::servoanisocor}
In most cases, bandwidth and anisoplanatism are the main contributors of the error breakdown of a SCAO system. Even if the correlation between anisoplanatism and bandwidth errors appears to be low, it is well known that it strongly depends on wind conditions. Under the frozen-flow assumption \citep{Taylor1938}, temporal fluctuations of the turbulence can be expressed as spatial fluctuations \citep{Greenwood1977}. Thus, we cannot neglect this correlation, especially in layer-by-layer error modeling as in \citep{Jolissaint2006}.

To illustrate this purpose and prepare further model validation, we ran 60 ROKET simulations with the same conditions as described in Section~\ref{sec:cases}, but with only a single layer at 5 km altitude and various wind conditions: a windspeed between 5 and 20 m/s, and a wind direction between 0$^\circ$ and 180$^\circ$.
We compare then the PSF$_R$ computed by ROKET, and the PSF$_I$ computed by neglecting anisoplanatism and bandwidth correlation. The comparison is presented in terms of SR and EE at $\pm 5 \frac{\lambda}{D}$.
Table~\ref{table::cor_results} summarizes the results obtained for each wind direction in terms of relative error on the SR and the EE.
\begin{table*}[!htb]
\centering
\begin{tabular}{c c c c c c }
\hline
\hline
\multirow{2}{*}{\textbf{Wind direction}}&\multirow{2}{*}{\textbf{Aniso/bandwidth correlation}}&\multicolumn{2}{c}{\textbf{SR relative error}} & \multicolumn{2}{c}{\textbf{EE relative error}} \\ 
& & mean & max & mean & max \\
\hline
0 & 0.93 & 46 \% & 84\% & 12 \% & 14 \% \\

45 & 0.63 & 26 \% & 47\% & 7 \% & 8 \% \\

90 & 0.01 & 0.4 \% & 2\% & 0.3 \% & 0.4 \% \\

135 & -0.66 & 24 \% & 38\% & 8 \% & 10 \% \\

180 & -0.95 & 31 \% & 48\% & 12 \% & 14 \% \\
\hline
\end{tabular}
\caption{\label{table::cor_results} Mean and maximum relative error between PSF$_I$ compared to PSF$_C$ over the 60 simulation runs, in terms of SR and EE in $\pm 5 \frac{\lambda}{D}$ for five values of the wind direction.}
\end{table*}
As expected, the anisoplanatism and bandwidth correlation depends on the wind direction. It is strongest when the wind is aligned
with the off-axis GS direction. Neglecting it in this case leads to huge errors in the PSF estimate. Conversely, the correlation is negligible when the wind direction is orthogonal to that of the GS.

%%%%%%%%%%%%%%%%%%%%%%%%%%%%%%%%%%%%%%%%%%%%%%%%%%%%
%%%%%%                                                                  Bandwidth & anisoplanatism errors                                  %%%%
%%%%%%%%%%%%%%%%%%%%%%%%%%%%%%%%%%%%%%%%%%%%%%%%%%%%

\section{Bandwidth and anisoplanatism models}
\label{sec::Model}
This work prepares some tools for PSF reconstruction based on the telemetry data of a real system. We propose an analytical model of the covariance matrix between bandwidth and anisoplanatism  to be retrieved from these telemetry and turbulence profiling tools.
Similar models already exist that aim at PSF computation \cite{Jolissaint2006, Neichel2008}. In this section, we propose a different approach that does not rely on a Fourier analysis, but on an expression of covariance matrices of the anisoplanatism and bandwidth errors, including their possible correlation, directly in the DM space. Applications and limitations of this approach are discussed.

\subsection{Definitions}
We define the bandwidth error as the temporal error due to the delay between the time when the wavefront is measured and the time when this command is applied on the DM. Then, we write the bandwidth error $\epsilon_b$ as a phase difference in the direction of the target between two moments that are separated by some delay $\tau'$:
\begin{equation}
\label{eq::bandwidth_def}
\epsilon_{b}=\Phi_{\parallel}(\mathbf{x},\mathbf{0},t+\tau') - \Phi_{\parallel}(\mathbf{x},\mathbf{0},t)
.\end{equation}
\label{sec::loop_gain}
We find the value of $\tau'$ assuming that we can approximate the rejection transfer function of the AO loop for one-frame delay $ H(p) = \frac{1}{1+g\frac{F_s}{p}e^{-\tau p}} $ \citep{Demerle1994} by the transfer function of substraction with pure delay $ H'(p) = 1 - e^{-\tau' p}$. The variable $p$ is the Laplace variable, $F_s$ is the sampling frequency of the AO system, and $\tau = 1/F_s$ is the sampling period. Considering the Taylor series of both expressions around $p=0$ at first order, they will match for the value 
\begin{equation}
\tau' = \frac{\tau}{g} \ .
\end{equation}
The validity of this assumption is discuss later.
In a real AO system, we should include the filtering by the exposure time on the WFS and by the blocker of the DM command. In any case, however, our pure delay identification procedure can still be applied. 

We define the anisoplanatism error as the phase difference between wavefront sensor and the target directions:
\begin{equation}
\label{eq::aniso_def}
\epsilon_{a}=\Phi_{\parallel}(\mathbf{x},\bm{\theta},t) - \Phi_{\parallel}(\mathbf{x},\mathbf{0},t)
,\end{equation}
where $\bm{\theta}$ is the angular separation between the WFS guide star and the target.

Now, we assume the case of a phase in a single layer. The multi-layers case will come easily under the frozen flow assumption. This will allow us to transform angular coordinates of observation into spatial coordinates at the layer surface, knowing the turbulent layer altitude $h$. We also assume a frozen flow hypothesis, which allows us to turn time into space using the wind velocity vector $\mathbf{v}$.
Eqs.~(\ref{eq::aniso_def}) and~(\ref{eq::bandwidth_def}) can both be rewritten in the target direction:
\begin{eqnarray}
\epsilon_{a}&=&\Phi_{\parallel}(\mathbf{x}+h\bm{\theta}) - \Phi_{\parallel}(\mathbf{x}) \\
\epsilon_{b}&=&\Phi_{\parallel}(\mathbf{x}-\mathbf{v}\tau') - \Phi_{\parallel}(\mathbf{x})
\label{eq::bandwidth_eq}
.\end{eqnarray}

We emphasize that in both definitions, we consider only the phase $\Phi_{\parallel}$ that belongs to the DM modal space. 

\subsection{Covariance matrix model}
From the previous definitions, we wish to derive an expression of the covariance matrix $C_{bb}$ of the bandwidth error, of the covariance matrix $C_{aa}$ of the anisoplanatism error, and of the covariance matrix $C_{ba}$ between these errors. We first compute these covariance matrices on the DM actuators.
Components $(i,j)$ of the covariance matrix $C_{ba}$ can be written as
\begin{equation}
C_{ba}(i,j) = \langle (\Phi_{\parallel}(\mathbf{x_i}+h\bm{\theta}) - \Phi_{\parallel}(\mathbf{x_i}))(\Phi_{\parallel}(\mathbf{x_j}-\mathbf{v}\tau') - \Phi_{\parallel}(\mathbf{x_j})) \rangle
,\end{equation}
where $\mathbf{x_i}$ and $\mathbf{x_j}$ are the position vectors in the pupil of the DM actuator number $i$ and $j$, respectively.
Using the identity
\begin{equation}
(A - a)(B - b) = \frac{1}{2} \left( -(A-B)^2 + (A-b)^2  \right.  \\
                                                                        \left.  + (a-B)^2 - (a-b)^2 \right)
\end{equation}
leads to
\begin{multline}
\label{eq::full_expres}
C_{ba}(i,j) = \frac{1}{2} \left(  - \langle (\Phi_{\parallel}(\mathbf{x_i}+h\bm{\theta}) - \Phi_{\parallel}(\mathbf{x_j}-\mathbf{v}\tau'))^2 \rangle \right. \\ 
+ \langle (\Phi_{\parallel}(\mathbf{x_i}+h\bm{\theta}) - \Phi_{\parallel}(\mathbf{x_j}))^2 \rangle  + \langle (\Phi_{\parallel}(\mathbf{x_i}) - \Phi_{\parallel}(\mathbf{x_j}-\mathbf{v}\tau'))^2\rangle \\
\left. - \langle (\Phi_{\parallel}(\mathbf{x_i}) -  \Phi_{\parallel}(\mathbf{x_j}))^2\rangle \right)
.\end{multline}
We introduce $D_{\phi}^{low}(\mathbf{r})$, the structure function of the turbulent phase restricted to the spatial frequencies of the DM space. We can write it as
\begin{equation}
D_{\phi}^{low}(\mathbf{r}) = \langle (\Phi_{\parallel}(\mathbf{x}+\mathbf{r}) - \Phi_{\parallel}(\mathbf{x}))^2 \rangle
.\end{equation}
Then, considering the separation vector $\mathbf{x_{ij}} = \mathbf{x_j} - \mathbf{x_i}$ between actuator $i$ and $j$, Eq.~(\ref{eq::full_expres}) can be written as
\begin{multline}
\label{eq::Cba}
C_{ba}(i,j) = \frac{1}{2} \left( -D_ {\phi}^{low}(\mathbf{x_{ij}}-h\bm{\theta}-\mathbf{v}\tau') + D_ {\phi}^{low}(\mathbf{x_{ij}}-h\bm{\theta}) \right. \\
\left.+ D_ {\phi}^{low}(\mathbf{x_{ij}}-\mathbf{v}\tau') - D_ {\phi}^{low}(\mathbf{x_{ij}})\right) 
.\end{multline}

The same approach stands for covariance matrices $C_{aa}$ and $C_{bb}$ and leads to the very similar expressions
\begin{eqnarray}
\label{eq::Cae}
C_{aa}(i,j) = \frac{1}{2} \left( D_ {\phi}^{low}(\mathbf{x_{ij}}+h\bm{\theta}) + D_ {\phi}^{low}(\mathbf{x_{ij}}-h\bm{\theta}) - 2 D_ {\phi}^{low}(\mathbf{x_{ij}})\right) \\
\label{eq::Cbe}
C_{bb}(i,j) = \frac{1}{2} \left( D_ {\phi}^{low}(\mathbf{x_{ij}}+\mathbf{v}\tau') + D_ {\phi}^{low}(\mathbf{x_{ij}}-\mathbf{v}\tau') - 2 D_ {\phi}^{low}(\mathbf{x_{ij}})\right) 
.\end{eqnarray}

These expressions are simple and fully analytical as we are able to compute the structure function $D_{\phi}^{low}(\mathbf{r})$. We define $f_c$, and the Nyquist spatial frequency of the DM equals to $\frac{1}{2d,}$ where $d$ is the actuator pitch.
Using a Fourier approach, this structure function limited to the spatial frequencies lower than $f_c$ can be written as
\begin{multline}
D_{\phi}^{low}(r) = D_{\phi}(r,L_0) - 2 (2\pi )^{8/3}0.023\left(\frac{r}{r_0}\right)^{5/3} \\
\left(1.11833-\int_0^{2\pi f_cr}u^{-8/3}(1-J_0(u)) \, \mathrm{d}u \right)
.\end{multline}
$\text{The }D_{\phi}(r,L_0)$ expression can be found in \citet{Conan2000}. This structure function is an approximation that considers circular symmetry and not the actual actuator pattern. The detailed calculation and computation methods are reported in Appendix~\ref{app::calcul_dphi}.
Then, we obtain the full contribution of the anisoplanatism and bandwidth errors by adding these covariance matrices:
\begin{equation}
\label{eq::Cee_phi}
C_{ee}^{\Phi} = C_{bb} + C_{aa} + C_{ba} + C_{ba}^t
.\end{equation}
The covariance matrix of a complex, multi-layer profile can be obtained by summing all the single-layer matrices together. 

Finally, we note with $C_{ee}$ the covariance matrix expressed in the actuator space and filtered from piston and tip-tilt modes. The removed tip-tilt component is projected on the tip-tilt mirror actuators.

This model only needs a few parameters from the atmospheric turbulence ($L_0$, $r_0$, turbulent profile, and wind) and from the AO system (loop frequency, actuator positions, and  guide star position).

\subsection{Model result}
\label{sec::model_methods}
To assess the validity of our model, we generated PSFs following two methods. The first method used the error buffers from ROKET of bandwidth and anisoplanatism, wich are summed together and converted into a covariance matrix that will feed the PSF reconstruction algorithm. The second method used Eq.~(\ref{eq::Cee_phi}) to compute the $C_{ee}$ , which also feeds the PSF reconstruction algorithm.

\begin{figure}[!htbp]
\centering
\resizebox{\hsize}{!}{\subfigure{\includegraphics{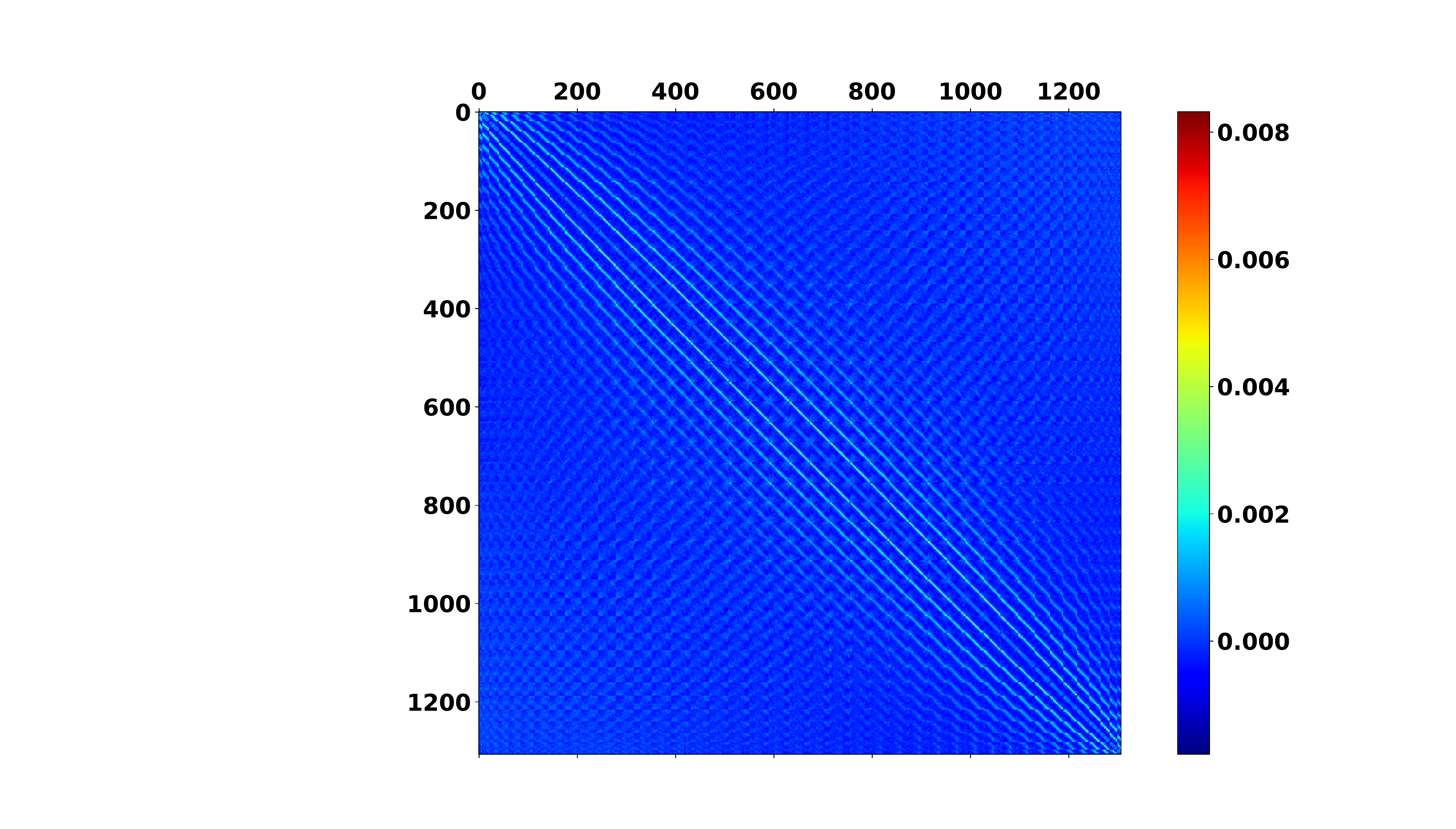}}}
\resizebox{\hsize}{!}{\subfigure{\includegraphics{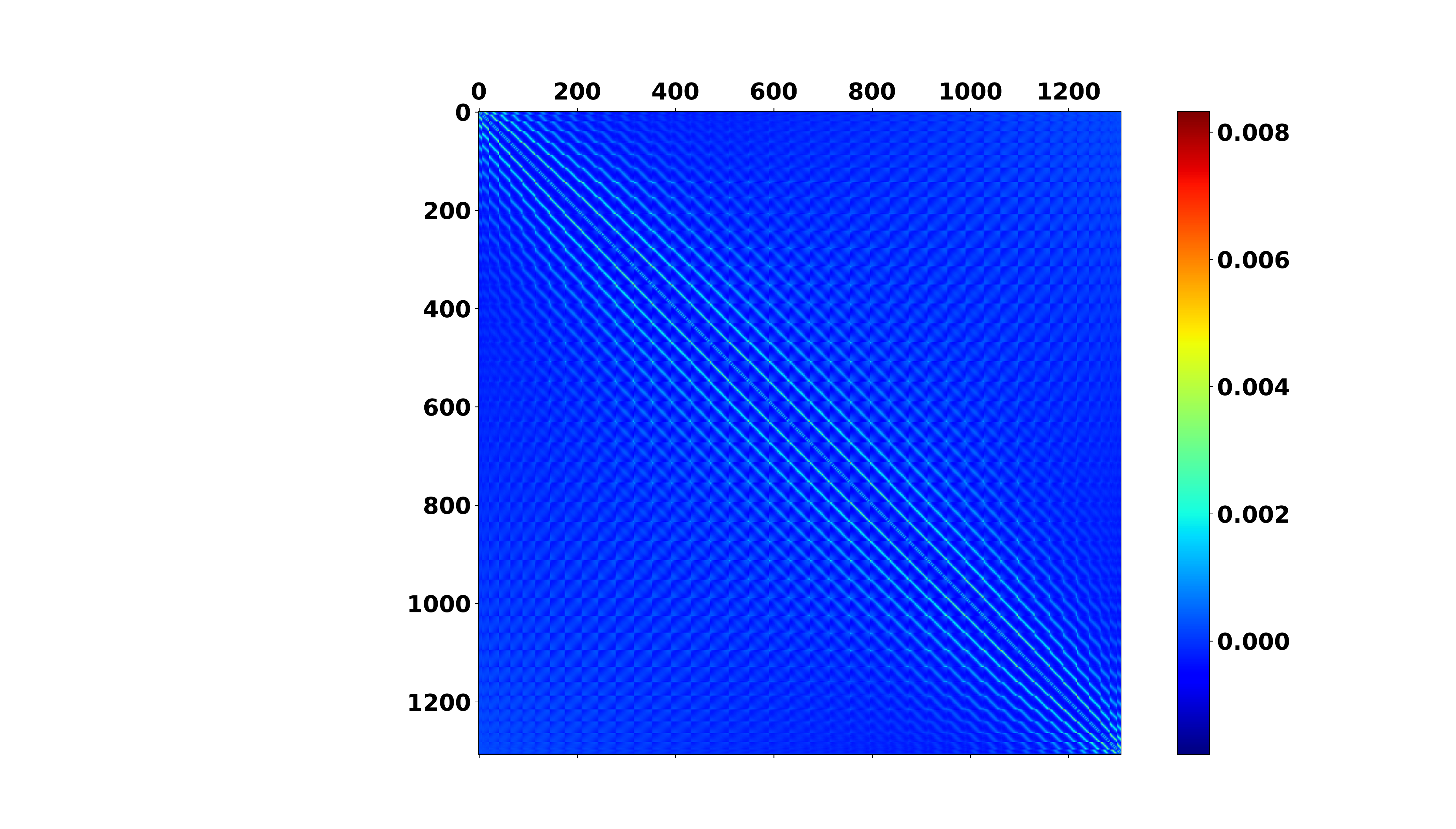}}}
\caption{\label{fig::covmats} Covariance matrices of the global contribution of the anisoplanatism and bandwidth errors. Both matrix diagonals have been nullified to emphasized side structures. Top: Covariance matrix computed from ROKET. Bottom: Covariance matrix model $C_{ee.}$}
\end{figure}

\begin{figure}[!htbp]
\centering
\resizebox{\hsize}{!}{\includegraphics[trim={8cm 0 0 0},clip]{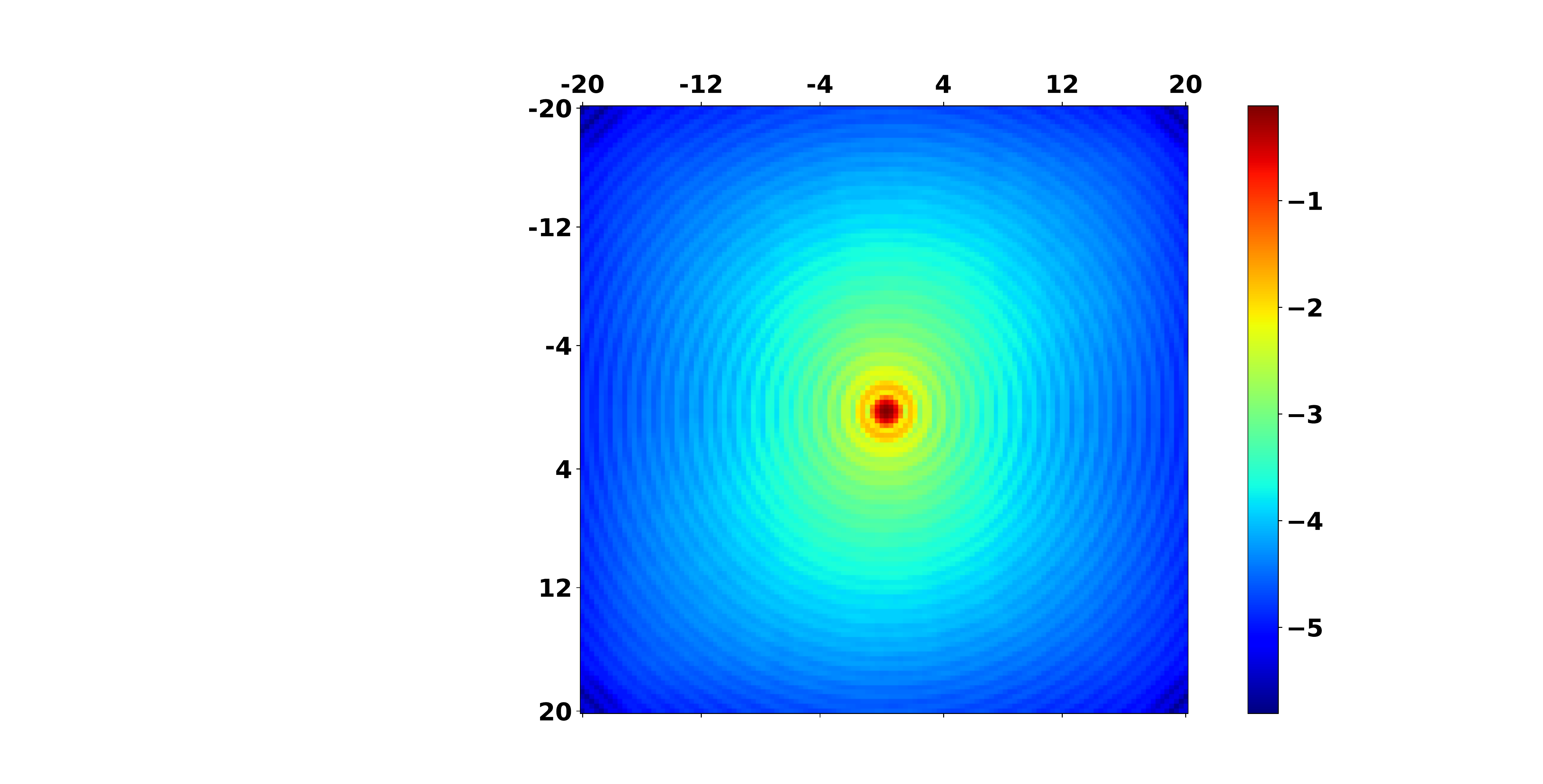}}
\resizebox{\hsize}{!}{\includegraphics[trim={8cm 0 0 0},clip]{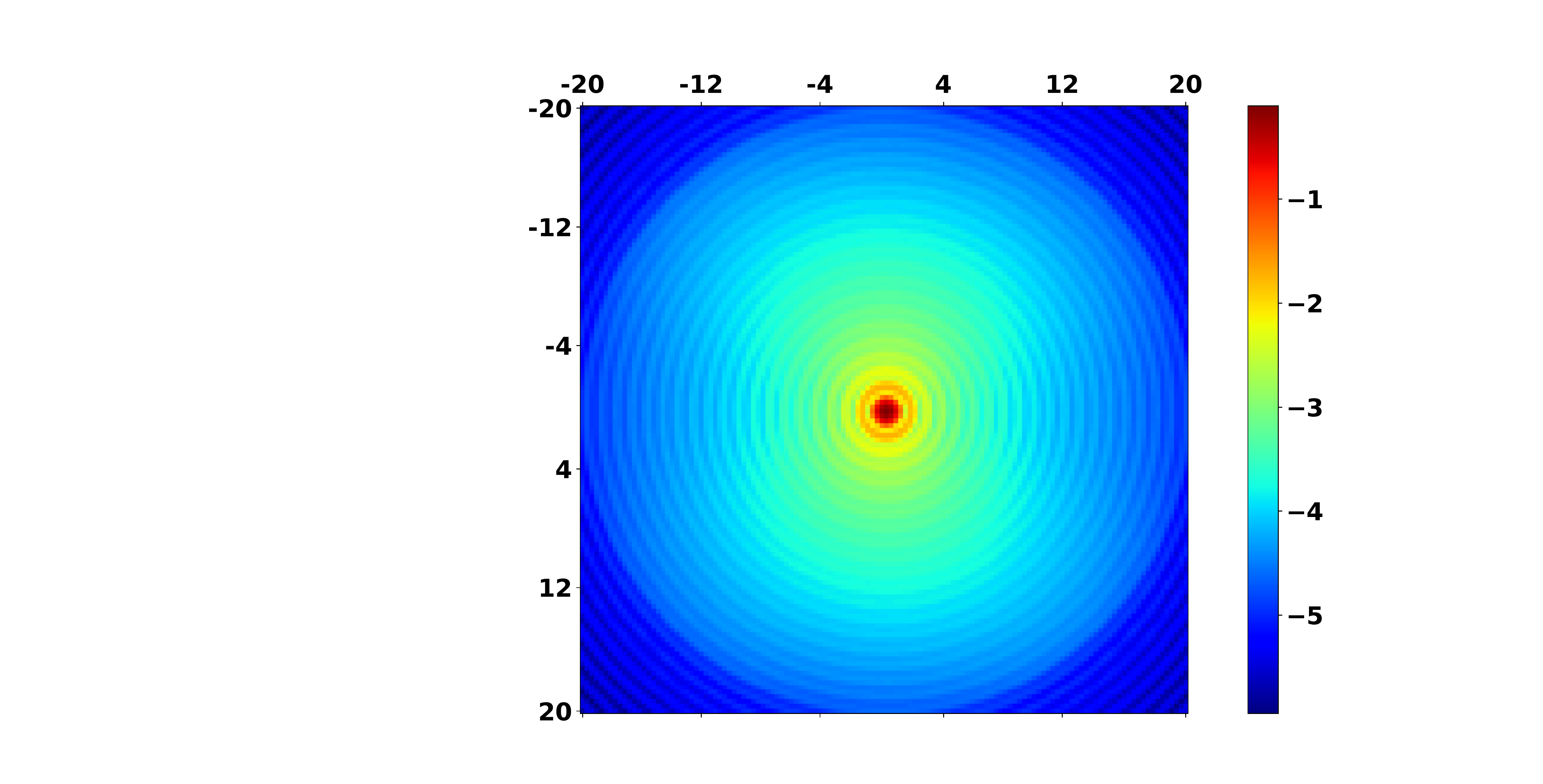}}
\resizebox{\hsize}{!}{\includegraphics[trim={8cm 0 0 0},clip]{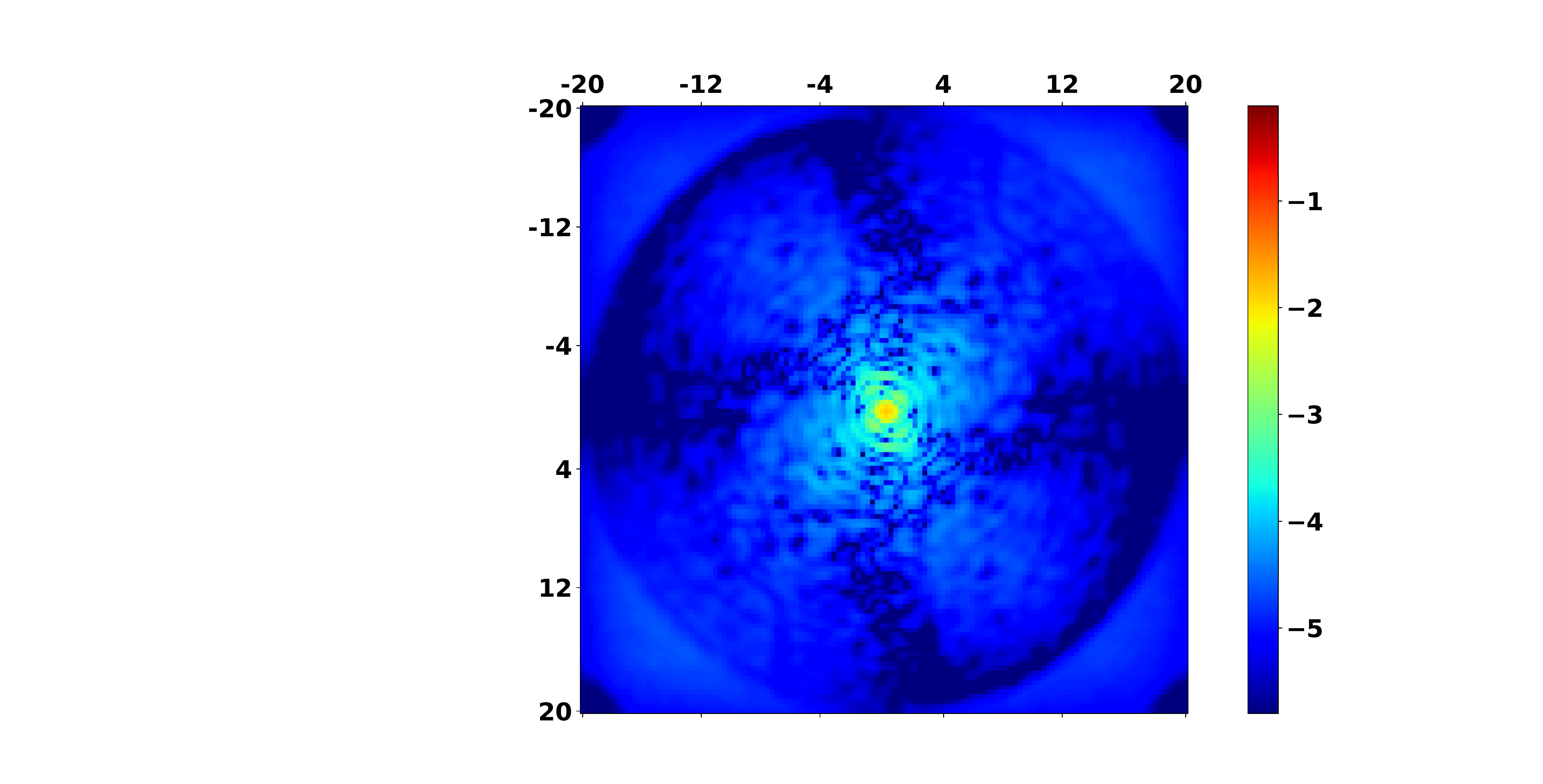}}
\caption{\label{fig::PSFs} Top: PSF obtained from ROKET buffers of anisoplanatism and bandwidth errors. Middle: PSF obtained from $C_{\epsilon \epsilon}$. Bottom: Absolute difference between the two PSFs. The actuator pattern is a square array. All figures are in log scale, and axes are expressed in units of $\frac{\lambda}{D}$.}
\end{figure}

\begin{figure}[!htbp]
\centering
\resizebox{\hsize}{!}{\includegraphics{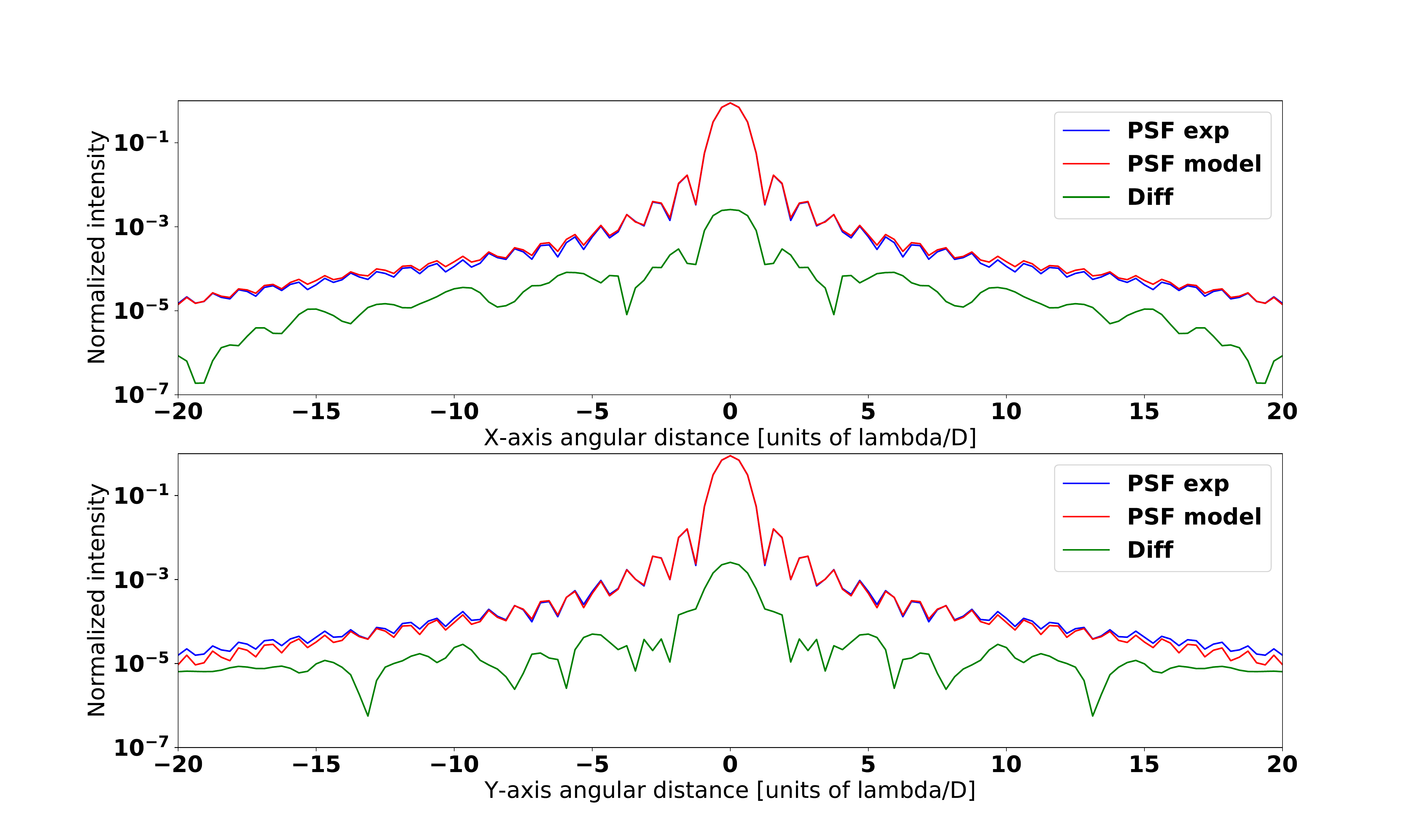}}
\caption{\label{fig::PSFs_cuts} Cuts of the PSFs in log scale. Top: Along the X-axis. Bottom: Along the Y-axis. Blue curves: PSF reconstructed from ROKET buffers. Red curves: PSF reconstructed from the model. Green curves: Absolute difference between the
two PSFs.}
\end{figure}

Figure~\ref{fig::covmats} shows both covariance matrices expressed in the DM actuators space. We note that the model reproduces the features of the covariance matrix well. Fig.~\ref{fig::PSFs} shows the PSFs obtained through the two methods and their difference. Cuts of these PSFs along the X- and Y-axes are shown in Fig.~\ref{fig::PSFs_cuts}.


We note a good accuracy in the PSF reconstruction up to $5 \frac{\lambda}{D}$, with a maximum intensity at 0.78 compared to 0.77, and an EE at $\pm 5 \frac{\lambda}{D}$ that equals 87.5 \% compared to 86.8 \%. Then, the PSF estimation at distances farther than $10 \frac{\lambda}{D}$ is less accurate.

%%%%%%%%%%%%%%%%%%%%%%%%%%%%%%%%%%%%%%%%%%%%%%%%%%%%
%%%%%%                                                                  LIMITATIONS                              %%%%
%%%%%%%%%%%%%%%%%%%%%%%%%%%%%%%%%%%%%%%%%%%%%%%%%%%%

\subsection{Model limitations}
\label{sec::limitations}
In this section, we apply the model we developed to the 60 simulations used in Section~\ref{sec::servoanisocor}. The results obtained highlight some limitations of the model that we explain below.

\begin{figure}[!htbp]
\centering
\resizebox{\hsize}{!}{\includegraphics{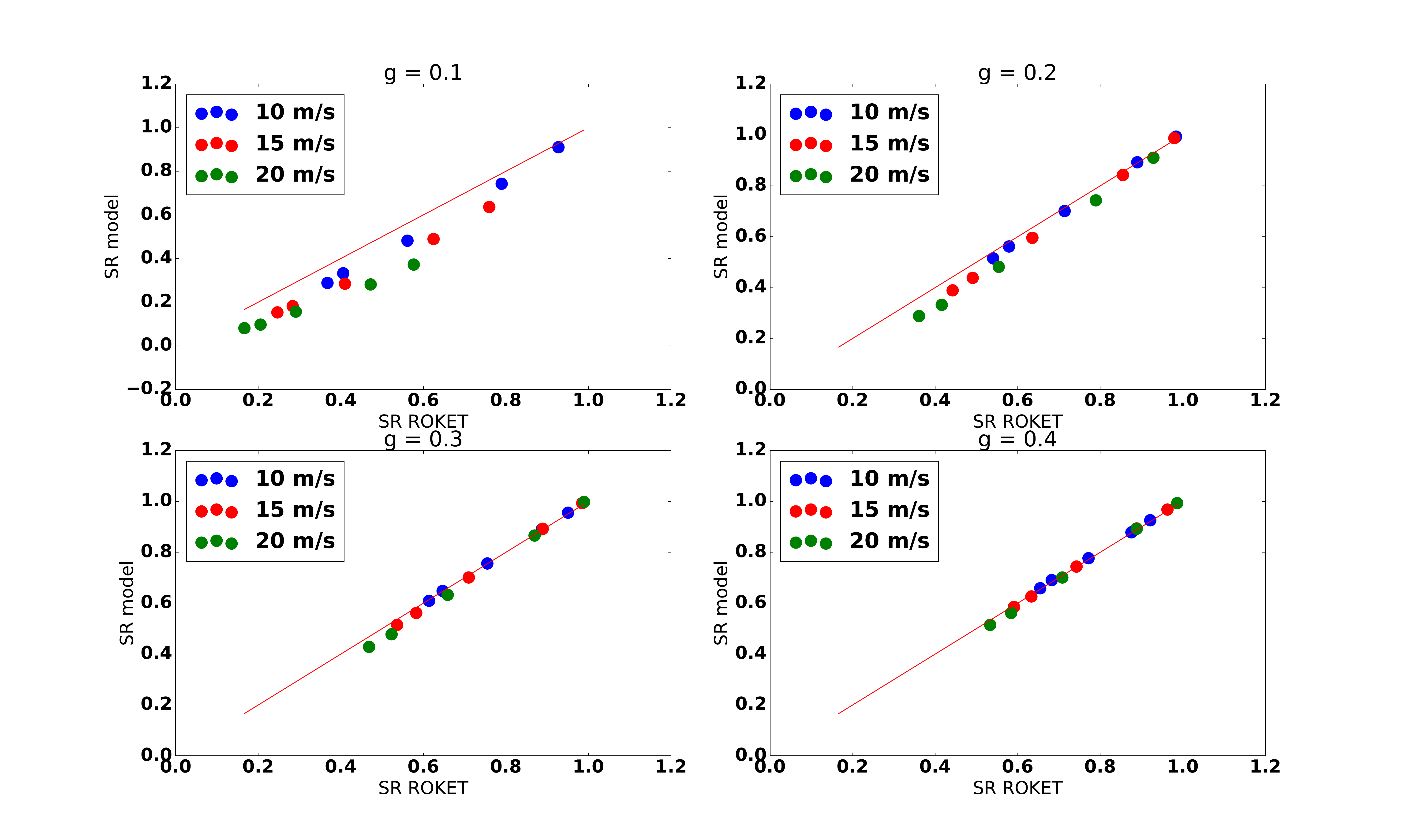}}
\caption{\label{fig::SRprecision} SR of PSF reconstructed from the model vs. SR of PSF reconstructed from ROKET buffers. Top left: Loop gain of 0.1. Top right: Loop gain of 0.2. Bottom left: Loop gain of 0.3. Bottom right: Loop gain of 0.4. Blue points: Wind speed of 10 m/s. Green: 15 m/s. Red: 20 m/s. The red line is $y=x$.}
\end{figure}

\begin{figure}[!htbp]
\resizebox{\hsize}{!}{\includegraphics{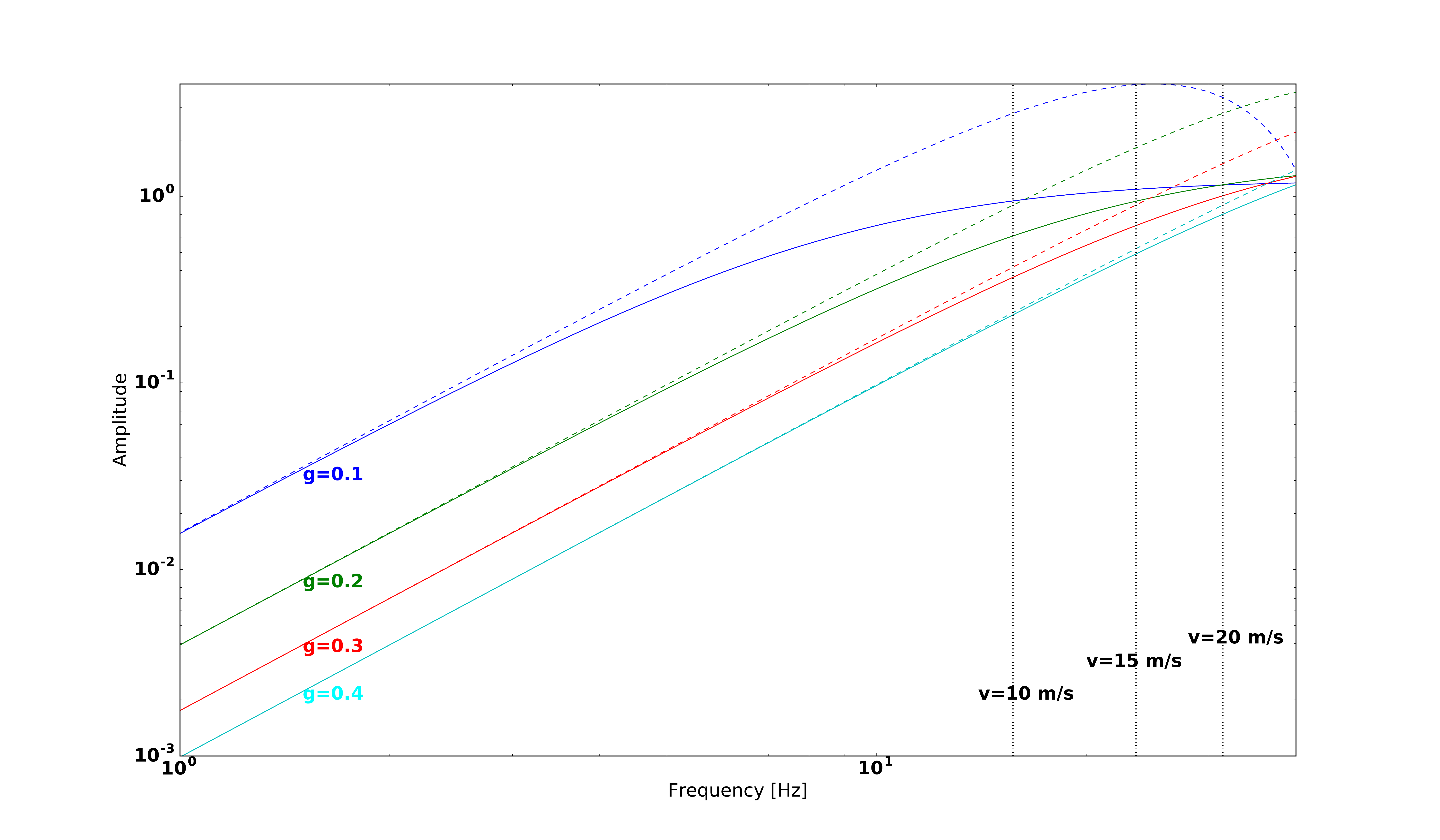}}
\caption{\label{fig::transfer_functions} Square modulus of the rejection transfer functionand the delay transfer function. Solid line: rejection transfer function, dashed line: delay transfer function, black dot lines: turbulence cut-off frequency depending on the wind speed}
\end{figure}

We now apply the same methods as we used in Section~\ref{sec::model_methods} to compute PSFs for the same 60 simulation cases.
Figure~\ref{fig::SRprecision} shows the SR obtained from a PSF reconstructed from $C_{\epsilon \epsilon}$ versus the one obtained from PSF reconstructed from ROKET buffers of anisoplanatism and bandwidth errors.
Globally, the faster the wind and the lower the loop gain, the less efficient the model. This is explained by the approximation we made in Section~\ref{sec::loop_gain} concerning the rejection transfer function. This assumption is made to obtain a simple model that could be computed fast and accurate in most cases, as shown in Section~\ref{sec::Model}. However, the validity of this assumption depends on the loop gain, the loop frequency, and the wind speed. Figure~\ref{fig::transfer_functions} compares the modulus of the rejection transfer function and the modulus of the delay transfer function for various gains.
The higher the gain, the closer the transfer functions, and so the more valid the assumption. Moreover, the approximation appears
to be better for low frequencies than for high frequencies. As the cut-off frequency of the turbulence is given by $f_c \approx 0.3 \frac{v}{d}$ \citep{Conan1995}, where $d$ is the subaperture diameter, the approximation, and so the model that we have developed, is better for low wind speed. The results obtained with Fig.~\ref{fig::SRprecision} confirm this behavior. 

\begin{figure}[!htbp]
\resizebox{\hsize}{!}{\includegraphics{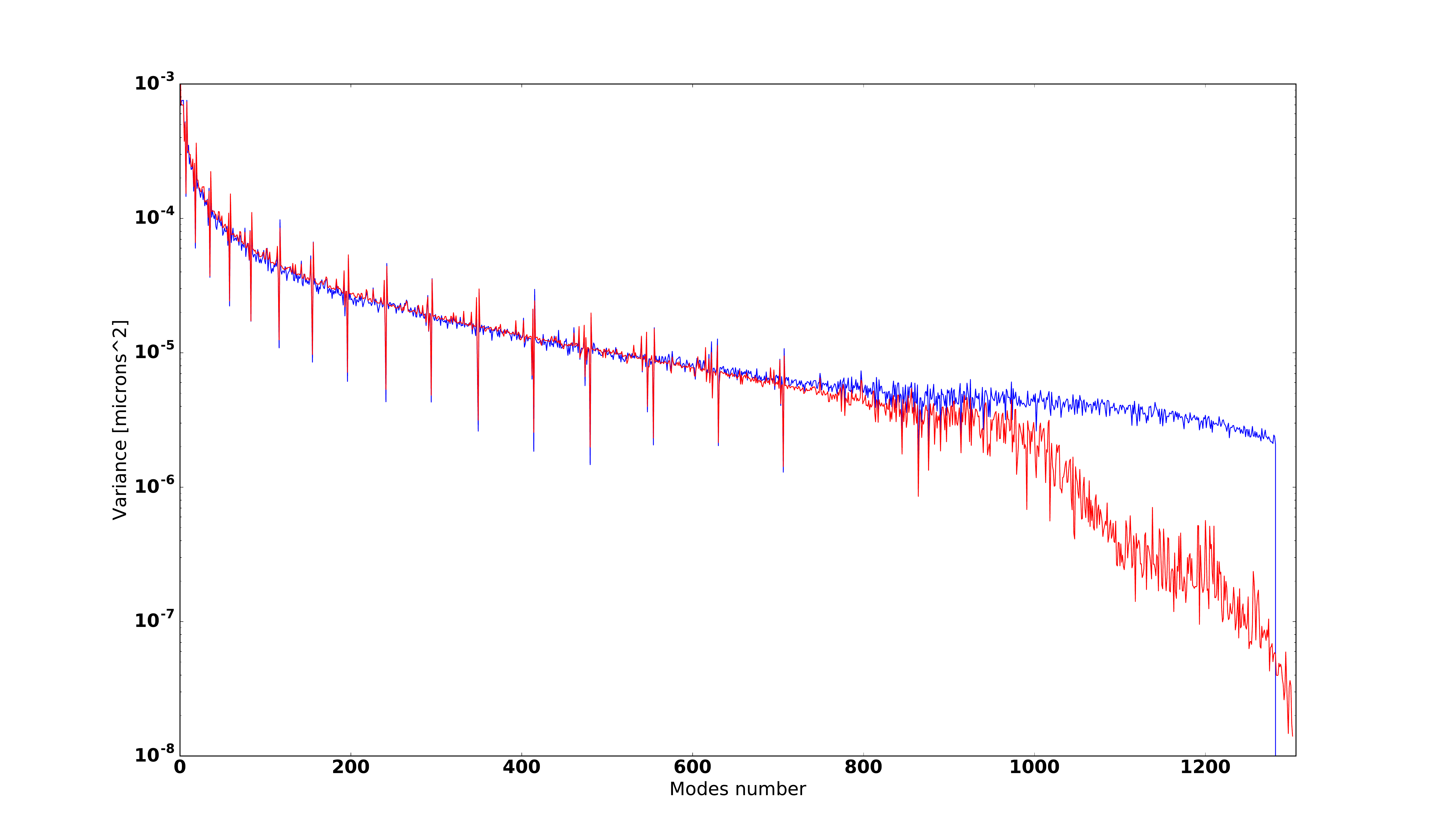}}
\caption{\label{fig::cov_on_modes} Diagonal of the error covariance matrix. Blue curve: Covariance computed from ROKET. Red curve: Covariance computed from the model.}
\end{figure}

It also leads to a poor estimation of the high-frequency components of the error. Figure~\ref{fig::cov_on_modes} shows the diagonal of the covariance matrix $C_{\epsilon \epsilon}$ in the modal basis space, compared to the covariance matrix computed from ROKET.
The model is accurate for modes from 0 to 800, then the covariance for higher modes is underestimated.

However, the results obtained with the model encourage us to keep the assumptions we made. For most cases, the accuracy is relevant and the computing performance obtained with this model is promising.

\subsection{Model applications}
As described above, models for anisoplanatism and bandwidth errors are available that use a power spectral density analysis of the AO residual phase \cite{Jolissaint2006}. The model that we propose is based on a slightly different approach that will be extended in future works to model the other error breakdown contributors. It will provide another method for estimating the PSF of an AO system. It will allow quick first-order AO performance evaluation even at the ELT scale.

Moreover, this approach can easily take into account the DM actuator geometry such as the hexagonal pattern and the shape of the influence functions within the limit of a modal basis that is restricted to a circular frequency domain. It could also be used in a PSF reconstruction algorithm.
It might also be useful in practical cases since it is based on an estimation of covariance matrices in the DM actuator space, which is a space available from AO loop telemetry. For example, a possible application could be the identification of turbulent parameters from AO loop data using a method similar to the one developed by \citet{Vidal2010} based on a fitting algorithm that minimizes the difference between the modeled covariance matrix with the matrix computed from telemetry data. Thus, retrieving the turbulence parameters from an SCAO system will be possible.

%%%%%%%%%%%%%%%%%%%%%%%%%%%%%%%%%%%%%%%%%%%%%%%%%%%%
%%%%%%                           COMPUTING PERFORMANCE                           %%%%
%%%%%%%%%%%%%%%%%%%%%%%%%%%%%%%%%%%%%%%%%%%%%%%%%%%%

\section{Computing performance}
\label{sec::computing_performance}
The simulations presented in this paper have been run on a Nvidia DGX-1 server, with Dual 20-core Intel Xeon E5-2698 v4 @ 2.2 Ghz and 8 Tesla V100-SXM2 GPUs.
The GPU acceleration enables COMPASS to run the 12-layer simulation, without the error estimation feature, at approximately 330 frames per second. Performing the error breakdown estimation, the computation speed decreases by a factor 5 at 60 frames per second. 

At the ELT scale, COMPASS is able to run a 35-layer SCAO simulation at 73 frames per second, and the error estimation is performed at 15 frames per second. 

We have also developed a multi-GPU module to perform PSF reconstruction using the algorithm proposed by \citet{Gendron2006}. This implementation allows reconstructing a 4 096 by 4 096 ELT PSF on 5 000 modes in 35 seconds on two GPUs.

The covariance matrix model described in Section~\ref{sec::Model} is highly parallelizable as each element of the matrix can be computed independently of the other elements. Hence, we have developed a GPU module dedicated to the computation of those matrices that lead to a computation of an ELT scale covariance matrix 5 000 by 5 000 over 35 layers in half a second.
%%%%%%%%%%%%%%%%%%%%%%%%%%%%%%%%%%%%%%%%%%%%%%%%%%%%
%%%%%%                     CONCLUSION                                                             %%%%
%%%%%%%%%%%%%%%%%%%%%%%%%%%%%%%%%%%%%%%%%%%%%%%%%%%%

\section{Conclusion and perspective}
We have developed a numerical tool, called ROKET, which is included in the COMPASS AO simulation code. It produces a comprehensive breakdown of the wavefront error for an AO loop. The estimated contributors are the temporal error, the anisoplanatism, the aliasing, the WF measurement deviation including centroid gain, the fitting, and the filtered mode error. The tool allows us to evaluate not only the variances, but the possible correlations between the different contributors to the error breakdown. 

Our results show that the error breakdown estimation is accurate and fast. The PSF, reconstructed from a ROKET error breakdown, is very similar to the empirical long-exposure PSF computed during the simulation run by COMPASS. The computed error breakdowns show that correlations exist between WF measurement deviation, noise, aliasing, temporal error, and anisoplanatism. We focused on the bandwidth and anisoplanatism errors, which are two main contributors in off-axis WFS. Their correlation often cannot be neglected.

We have developed an analytical model of these two main contributors and their correlation. It allows us to compute the resulting error covariance matrix that can be directly used by classical PSF reconstruction algorithms. Our results show that this model is accurate, even if the accuracy reached for the highest order controlled modes is lower than the one for low-order modes.

This model only requires a few parameters of the AO system and of the atmospheric conditions to compute the covariance matrix. The novelty compared to other analytical models is that it is expressed directly on the DM actuator space, allowing then convenient implementation for post-facto PSF reconstruction, for instance. In addition, the computing efficiency can be greatly improved by a GPU implementation, which allows easily handling an ELT scale.

Future works will use ROKET as a reference tool to develop the same type of models for the other error breakdown contrib- utors. With such models, fast and accurate PSF estimation of an AO system will be possible, even at the ELT scale. We plan to use ROKET for the validation by simulation of a turbulent pro- file identification tool from SCAO telemetry data. Another step will be to upscale ROKET and the covariance model for MCAO or MOAO systems.

%%%%%%%%%%%%%%%%%%%%%%%%%%%%%%%%%%%%%%%%%%%%%%%%%%%%
%%%%%%                                                                  ACKNOWLEDGEMENTS & BIBLIO                                    %%%%
%%%%%%%%%%%%%%%%%%%%%%%%%%%%%%%%%%%%%%%%%%%%%%%%%%%%
\begin{acknowledgements}
This work is sponsored through a grant from project \#671662, a.k.a. Green Flash, funded by European Commission under program H2020-EU.1.2.2 coordinated in H2020-FETHPC-2014.

We also wish to thank the anonymous referee for their meaningful comments on this article.
\end{acknowledgements}

\bibliographystyle{aa} % makes bibtex use spiebib.bst
\bibliography{biblio} % bibliography data in report.bib

\begin{appendix}
\section{Modal basis $\mathcal{B}_{tt}$ computation}
\label{app::modal_basis}
As specified in Section~\ref{modalBasis} , the error breakdown needs to be estimated on a modal basis with specific properties: 
\begin{itemize}
\item the modes span the full DM space
\item the modes are normalized ($\frac{1}{S} \int B_i^2 dS= 1 \mu m^2, \forall i$ with the integral defined over the pupil area)
\item they are orthogonal, in the sense that the scalar product over the pupil area is null ($\frac{1}{S} \int B_i B_j dS= 0$ for $i \neq j$)
\item the subspace of the modes commanded through the system control matrix is orthogonal (same scalar product as above) to the subspace of filtered ones
\item one of the filtered modes is constructed as the best DM least-squares fit to piston, so that all modes are orthogonal to the piston ($\forall i, \int B_i dS= 0$). This property is important for deriving phase variances that do not include any component along the piston term.
\item Two of these modes are pure tip-tilt and are associated with the tip-tilt mirror, while the tip-tilt that can be produced by the DM was suppressed.
\end{itemize}

To ensure that our modal basis $\mathcal{B}_{tt}$ spans the full DM space, it is computed from the influence functions of the DM. Let $IF$ the basis of influence functions of the DM, that
is, any phase produced by the DM can be decomposed as
\begin{equation}
\Phi_{DM} = \sum_{i=1}^{N_{actus}} a_i \, IF_i
.\end{equation}
Then, the dimensions of this basis are $N_{\Phi} \times N_{actus}$ , where $N_{\Phi}$ is the number of points in the pupil area and $N_{actus}$ is the number of DM actuators. The geometric covariance matrix $\Delta$ \citep{Gaffard1987} is then defined as
\begin{equation}
\Delta = IF^t . \, IF
\end{equation}
As $IF$ is a basis, $\Delta$ is invertible and $\Delta^{-1} \, . \, IF^t$ projects a phase onto the $IF$ basis.

Let $T_p$ be the matrix containing the phase corresponding to a piston and pure tip-tilt. The dimensions of this matrix are
therefore $ N_{\Phi} \times 3$. Using the projection matrix defined above, we can retrieve the coefficients $a_i$ on the basis $IF$ that fits the piston, tip-tilt modes, and store them in a matrix $\tau$:
\begin{equation}
\tau = \Delta^{-1} . \, IF^t . \, T_p
.\end{equation}
Our $\mathcal{B}_{tt}$ is to be orthogonal to the piston mode, and this includes pure tip-tilt modes. Then, we have to generate a set of generators $G$ from $IF$ that cannot produce those modes \citep{Gendron1995}: 
\begin{equation}
G = I - \tau\, . \,  \left( \tau^t . \, \Delta \, . \, \tau \right)^{-1} . \, \tau^t . \, \Delta
,\end{equation}
where $I$ is the identity matrix. Now, the diagonalization of $G$ provides a basis $B'$:
\begin{equation}
G^t. \, \Delta \, . \, G = B' \, . \, \lambda \, . \, B'^t
,\end{equation}
where $\lambda$ are the eigenvalues of $G^t. \, \Delta \, . \, G$ . After truncation of the three last columns of $B'$ corresponding to piston, tip-tilt modes, and normalization,
\begin{equation}
B = G \, . \, B' \, . \, \sqrt{\lambda}^{-1}
,\end{equation}
we obtain an orthonormal basis $B$ such that $B^t \, .\, \Delta \,.\, B = I$.

Finally, we have to add the pure tip-tilt modes in the basis $B$ to obtain the modal basis $\mathcal{B}_{tt}$,
\begin{equation}
\mathcal{B}_{tt} = \begin{pmatrix} & & & 0 & 0 \\ & B & & \vdots & \vdots \\ & & & 0 & 0 \\ 0 & \cdots & 0 & \frac{1}{\mu} & 0 \\ 0 & \cdots & 0 & 0 & \frac{1}{\mu} \end{pmatrix}
,\end{equation}
where $\mu$ are the eigenvalues of the geometric covariance matrix of the tip-tilt modes, corresponding to its diagonal as these two modes composed a basis.

\section{AO loop error with centroid gain}
\label{app::centroid_gain}
A centroid gain $\gamma$  will have an effect on the equations described in \cite{Ferreira2016} and Section~\ref{sec::ROKET}. To take it into account in ROKET, we have to rewrite the equations with a centroid gain. It can be modeled as a gain $\gamma$ on the WFS measurement, so that Eq.~(\ref{eq::WFS_measure}) becomes\begin{equation}
    \mathbf{w}_k = \gamma D \mathbf{a_{k,\theta}} + \gamma D \mathbf{v_{k-1}} + \gamma \mathbf{r}_k + \mathbf{n}_k + \mathbf{u}_k
.\end{equation}
In this expression, $\mathbf{u}_k$ is no longer the same vector as in Eq.~(\ref{eq::WFS_measure}) as it does not include the centroid gain error.

Then, the command vector $\mathbf{v}_k$ can be written as
\begin{equation}
\mathbf{v}_k = (1 - \gamma g R D) \mathbf{v}_{k-1} - \gamma g R D \mathbf{a}_{k,\theta} - \gamma g R \mathbf{r}_k - g R \mathbf{n}_k - g R \mathbf{u}_k
.\end{equation}
Finally, the residual error $\bm{\epsilon}_k$ becomes
\begin{multline}
\bm{\epsilon}_k = (1 - \gamma g R D) \bm{\epsilon}_{k-1} + (\mathbf{a}_k - \mathbf{a}_{k-1}) + \gamma g R D(\mathbf{a}_{k-1}-\mathbf{a}_{k-1,\theta}) \\
- \gamma g R \mathbf{r}_{k-1} - g R \mathbf{n}_{k-1} - g R \mathbf{u}_{k-1.}
\end{multline}

Hence, the equations for the estimation of the error breakdown contributors described in Section~\ref{section:contributors} become
\begin{eqnarray}
\label{eq:newCentroidGain}
\begin{aligned}
    \bm{\beta}_k &= (1 - \gamma g R D) \bm{\beta}_{k-1} + (\bm{a}_k - \bm{a}_{k-1}) \\
    \bm{\tau}_k &= (1 - \gamma g R D) \bm{\tau}_{k-1} + \gamma g  R D (\bm{a}_{k-1} - \bm{a}_{k-1,\theta}) \\
    \bm{\rho}_k &=(1 - \gamma g R D) \bm{\rho}_{k-1} - \gamma g R \bm{r}_{k-1} \\
    \bm{\eta}_k &= (1 - \gamma g R D) \bm{\eta}_{k-1} - g R \bm{n}_{k-1}\\
    \bm{\mu}_k &= (1 - \gamma g R D) \bm{\mu}_{k-1} - g R \bm{u}_{k-1}
\end{aligned}
.\end{eqnarray}

\section{Calculation of the low-frequency structure function $D_{\phi}^{low}$}
\label{app::calcul_dphi}
To compute a model for the covariance between anisoplanatism and bandwidth error as a sum of structure functions, we have to compute these functions. However, as we are searching for errors made on the DM command, we need to calculate the structure function that excludes spatial frequencies higher than the Nyquist frequency $f_c=\frac{1}{2d}$ of the DM.

Starting from the Kolmogorov spectrum,
\begin{equation}
W(k) = 0.023\, r_0^{-5/3}(k)^{-11/3}
,\end{equation}
the structure function that we search for can be written as
\begin{align}
D_{\phi}^{low}(r) &  =  2\int_0^{2\pi} \int_0^{f_c} W(k)(1-cos(2\pi k r) k dk d\theta \\
 & =  2\times\, 0.023 \, r_0^{-5/3} \nonumber \\ & \quad \times \int_0^{f_c} \int_0^{2\pi} k^{-8/3}(1-cos(2\pi kr cos(\theta))dk d\theta \\
 & =  2\times\, 0.023 \, r_0^{-5/3}\, \nonumber \\ & \quad \times 2\pi \int_0^{f_c} k^{-8/3}(1-J_0(2\pi k r))dk. 
\end{align}
With the variable change $u=2\pi k r$, we obtain
\begin{align}
D_{\phi}^{low}(r) &  =   2\times\, 0.023 \, r_0^{-5/3}\, \nonumber \\ & \quad \times 2\pi \int_0^{2\pi f_c r} u^{-8/3} (2\pi r)^{8/3}(1-J_0(u))\frac{du}{2\pi r} \\
& = 2\times\, 0.023 \, (2\pi)^{8/3}\left(\frac{r}{r_0}\right)^{5/3}\, \nonumber \\ & \quad \times \int_0^{2 \pi f_c r} u^{-8/3}(1-J_0(u))du.
\label{eq::dphi_low_kolmo}
\end{align}

However, this expression does not take into account the outer scale of the turbulence $L_0$. To include this, we note that
\begin{equation}
\int_0^{\infty} u^{-8/3}(1-J_0(u))du = 1.11833
.\end{equation}
From this result and Eq.~(\ref{eq::dphi_low_kolmo}), we derive an expression of the structure function limited to the frequencies higher than $f_c$:
\begin{multline}
D_{\phi}^{high} = 2\times\, 0.023 \, (2\pi)^{8/3}\left(\frac{r}{r_0}\right)^{5/3}\\
\left( 1.11833 - \int_0^{2 \pi f_c r} u^{-8/3}(1-J_0(u))du\right)
.\end{multline}
Moreover, we know that the effect of the outer scale on the spectrum is a saturation effect on low frequencies. Hence, the expression of $D_{\phi}^{high}$ is not affected since $f_c > \frac{1}{L_0}$. Then, we can obtain the final $D_{\phi}^{low}(r,L0)$ as the difference between the complete structure function $D_{\phi}(r,L_0)$ and $D_{\phi}^{high}(r)$:
\begin{equation}
D_{\phi}^{low}(r,L0) = D_{\phi}(r,L_0) - D_{\phi}^{high}(r)
.\end{equation}

\end{appendix}

\end{document}